\documentclass[preprint,aps,prd,floatfix,superscriptaddress,preprintnumbers,nofootinbib]{revtex4-1}

\pdfoutput=1

\usepackage{anyfontsize}
\usepackage{graphicx}
\usepackage{amssymb}
\usepackage{amsmath,amssymb}
\usepackage{color}
\usepackage{multirow}
\usepackage{wrapfig}

\newcommand{\beq}{\begin{equation}}
\newcommand{\eeq}{\end{equation}} 

\usepackage[bookmarks=false]{hyperref}

\allowdisplaybreaks

\begin{document}

\preprint{FERMILAB-PUB-14-310-T}

\title{\Large Fermion Hierarchy from Sfermion Anarchy}

\author{Wolfgang Altmannshofer}
\affiliation{Perimeter Institute for Theoretical Physics, Waterloo, ON, N2L 2Y5, Canada}
\author{Claudia Frugiuele}
\affiliation{Fermi National Accelerator Laboratory, P.O. Box 500, Batavia, IL 60510, USA}
\author{Roni Harnik}
\affiliation{Fermi National Accelerator Laboratory, P.O. Box 500, Batavia, IL 60510, USA}

\begin{abstract}
We present a framework to generate the hierarchical flavor structure of Standard Model quarks and leptons from loops of superpartners. The simplest model consists of the minimal supersymmetric standard model with tree level Yukawa couplings for the third generation only and anarchic squark and slepton mass matrices. Agreement with constraints from low energy flavor observables, in particular Kaon mixing, is obtained for supersymmetric particles with masses at the PeV scale or above. 
In our framework both the second and the first generation fermion masses are generated at 1-loop. Despite this, a novel mechanism generates a hierarchy among the first and second generations without imposing a symmetry or small parameters. A second-to-first generation mass ratio of order 100 is typical. The minimal supersymmetric standard model thus includes all the necessary ingredients to realize a fermion spectrum that is qualitatively  similar to observation, with hierarchical masses and mixing. The minimal framework produces only a few quantitative discrepancies with observation, most notably the muon mass is too low. 
We discuss simple modifications which resolve this and also  investigate the compatibility of our model with  gauge and Yukawa coupling Unification.
\end{abstract}

\maketitle

\tableofcontents

\section{Introduction}

The discovery of a Higgs boson~\cite{Aad:2012tfa, Chatrchyan:2012ufa} and the lack of evidence for  beyond the standard model (SM) physics has placed natural models of electro-weak symmetry breaking under stress. Most prominently, the framework of supersymmetry (SUSY) for addressing the hierarchy problem is looking more and more constrained~\cite{Craig:2013cxa}. As this situation continues, the possibility that the hierarchy problem, which after all is a theoretical one, is not addressed in the traditional ``natural'' sense should be considered. We are led to consider other problems facing the SM within frameworks in which the electro-weak scale is tuned. 

In this work we will consider a class of models that attempt to explain the flavor structure of fermion masses and mixings. There are many possible frameworks to get exponential fermion hierarchy from input parameters that are of order one. These include Froggatt-Nielsen models~\cite{Froggatt:1978nt,Leurer:1992wg,Leurer:1993gy}, fermion profiles in flat~\cite{ArkaniHamed:1999dc,Kaplan:2001ga} or warped~\cite{Grossman:1999ra,Gherghetta:2000qt} extra dimensions, and models with large anomalous dimensions~\cite{Nelson:2000sn}. In fact, within these frameworks it is quite generic to have hierarchies that are too large. 

Here we will focus on models of radiative fermion masses~\cite{Weinberg:1972ws}. Within this framework the heavy fermions receive their mass at the tree level and light fermions receive a mass through loops of new heavy particles. There are many realizations of this idea both in the context of supersymmetric models~\cite{Lahanas:1982et,Nanopoulos:1982zm,Masiero:1983ph, delAguila:1984qs, Banks:1987iu, Kagan:1987tpa,Kagan:1989fp,ArkaniHamed:1995fq,ArkaniHamed:1996zw,Hamzaoui:1998yy,Babu:1998tm,Borzumati:1999sp,DiazCruz:2000mn,Ferrandis:2004ng,Ferrandis:2004ri,Crivellin:2008mq,Crivellin:2010ty,Crivellin:2011sj,ArkaniHamed:2012gw,Baumgart:2014jya} and non-supersymmetric models~\cite{Barr:1979xt,Barbieri:1980tz,Ibanez:1981nw,Barbieri:1981yy,Balakrishna:1987qd,Balakrishna:1988ks,Balakrishna:1988bn,Ma:1988qc,Ma:1989tz,He:1989er,Berezhiani:1991ds,Dobrescu:2008sz,Hashimoto:2009xi,Graham:2009gr,Ibarra:2014fla,Joaquim:2014gba}.\footnote{For early attempts to explain the electron to muon mass ratio by a loop factor see also~\cite{Georgi:1972mc,Georgi:1972hy,Mohapatra:1974wk,Barr:1976bk,Barr:1978rv}.} The common ingredient of all these models is the inclusion of new states that mediate flavor violation. The variety of possibilities for quantum numbers and interactions leads to a plethora of models. An attractive feature of the framework of radiative fermion masses is that the typical mass hierarchy between fermions of adjacent generations is roughly a loop factor (with variations among models), in qualitative agreement with the observed spectra. A larger hierarchy is, of course, possible, but this requires additional small numbers or approximate symmetries. 

In our study we will use SUSY as a guiding principle for the particles and interactions that mediate flavor violation. Because our starting point is a tuned electro-weak scale, one could have expected that we can forget about SUSY altogether. However, as a unique extension of the space-time symmetries of the S-matrix in four dimensions, SUSY is well motivated at high scales, irrespective of its relevance for setting the electro-weak scale. Indeed, we will find that introducing SUSY at a high scale (a PeV or above) automatically includes all of the necessary ingredients to mediate flavor violation from the third generation fermions to those of the first and second families. A lower SUSY scale would also yield a similar fermion mass spectrum but at the cost of introducing unacceptably high rates of flavor changing neutral current processes. In this sense, when enumerating the benefits of supersymmetry, we are trading a few orders of magnitude in its solution to the hierarchy problem for an explanation of flavor. 

The minimal model we will consider is remarkably simple - it is a scaled up version of the minimal supersymmetric standard model (MSSM), with supersymmetry broken at or above a PeV and with tree level Yukawas just for the third generation. Supersymmetry breaking is assumed to be flavor anarchical and flavor violation is mediated from the third to the lighter generations at 1-loop. This simple setup will yield a spectrum that is qualitatively similar to observation, with a fully hierarchical spectrum and hierarchical mixings. The minimal setup will have a few quantitative problems which will be addressed subsequently.

Our framework differs from many other models of radiative fermion masses in that both the second and first generations receive a mass at the 1-loop level. One may worry that this generically does not give a hierarchical spectrum, but one in which the first and second generations are nearly degenerate.
However, this is not the case. As we shall see, thanks to a novel mechanism, the typical spectrum in our framework is fully hierarchical, typically giving a ratio of $1/100$ for first to second generation masses. Ratios of order one are not achievable. It should be noted that this small ratio is not the result of a small parameter or of a symmetry at the Lagrangian level. Instead it arises due to symmetry properties of the loop \emph{integrand} which mediates flavor violation, which are approximately preserved at the integral level. 

We will discuss the main mechanism qualitatively in section~\ref{sec:mechanism}.
In section~\ref{sec:spectrum} we will discuss the fermion spectrum in the minimal model quantitatively, both in the mass insertion approximation and by performing a numerical scan. We also discuss the constraints that vacuum stability place on the model.
An examination of quark mixing will be presented in section~\ref{sec:mixing}, where we will consider the cases of flavor conserving and flavor violating $A$-terms separately. 
In section~\ref{sec:mod} we will consider modifications of the minimal model that allow for milder hierarchies in the fermion spectrum, particularly raising the muon mass. 
In section~\ref{sec:unification} we comment on the compatibility of our frameworks with gauge and Yukawa coupling unification.
In section~\ref{sec:conclusion} we conclude.
Experimental values for the quark and lepton masses as well as CKM parameters are collected in appendix~\ref{app:masses}. A brief discussion of vacuum stability constraints in the presence of generic sfermion flavor mixing can be found in appendix~\ref{app:vac}. Finally, in appendices~\ref{app:flavor} and~\ref{app:higgs} we discuss constraints from low energy flavor observables and the Higgs mass.

\section{The Setup and a New Mechanism} \label{sec:mechanism}

We will begin by considering the most minimal setup to demonstrate the mechanism for radiative generation of flavor by sfermions. This simple setup will yield a fermion spectrum that is qualitatively similar to that observed in the SM, that of a hierarchical spectrum of fermion masses. 
In this section we will focus on the origin of the hierarchy. We will asses its success qualitatively in the next two section. This minimal setup has been considered briefly in the context of TeV scale SUSY~\cite{ArkaniHamed:1995fq,ArkaniHamed:1996zw}, only to be dismissed because of excessive flavor changing effects and/or the lack of naturalness. Giving-up on naturalness frees us to consider this model and some of its virtues in greater detail.

\subsection{The Minimal Setup} \label{sec:setup}

The particle content of the minimal setup includes just the particles of the MSSM with a high SUSY breaking scale and a tuned electro-weak scale (the spectrum is \emph{not} split as in split-SUSY~\cite{ArkaniHamed:2004fb, Giudice:2004tc,Wells:2004di}, distinguishing our model from~\cite{ArkaniHamed:2012gw,Baumgart:2014jya}). In the SUSY limit we assume only a single generation of fermions participates in a Yukawa interaction with the Higgs. We can thus rotate to a basis in which the Yukawa matrices take the form 
\begin{equation}
\label{eq:yukawa-tree}
Y_{u,d,e}\sim 
\begin{pmatrix} 
0 & 0 & 0 \cr
0 & 0 & 0 \cr
0 & 0 & 1 
\end{pmatrix}\,.
\end{equation}
This form can easily be arranged by assuming that matter interacts with the Higgs only through mixing with a single generation of heavy vector-like chiral superfields which ensures the Yukawa matrix is of rank 1 (see for example ~\cite{Kagan:1989fp, Dobrescu:2008sz}). This Yukawa interaction respects a global $U(2)^5$ flavor symmetry, with one $U(2)$ acting on the first two generations for each of the the $q,u,d,\ell$ and $e$ representations. 
In addition to generating a mass for the third generation, the couplings of equation~(\ref{eq:yukawa-tree}) induce non-holomorphic scalar trilinear terms of the form  
\begin{equation} \label{eq:mutrilinear}
\mathcal{L}_\text{SUSY} \supset \mu Y_u H_d^* \tilde q_3 \tilde u_3 + \mu Y_d H_u^* \tilde q_3 \tilde d_3 + \mu Y_e H_u^* \tilde \ell_3 \tilde e_3 + \mathrm{h.c.} ~,
\end{equation}
which obviously respect the same symmetry.

The mass scale of all superpartners, including scalars, gauginos and Higgsinos, will be set by a common SUSY breaking scale, $\tilde m$ which we will take to be at or above 1000 TeV. For now, we will assume that SUSY breaking trilinear terms such as $H_u\tilde q \tilde u$ are aligned in flavor space with the tree level Yukawa matrix of equation~(\ref{eq:yukawa-tree}), respecting the $U(2)^5$ flavor symmetry 
\begin{equation}
\mathcal{L}_\text{soft} \supset Y_u A_t H_u \tilde q_3 \tilde u_3 + Y_d A_b H_d \tilde q_3 \tilde d_3 + Y_e A_\tau H_d \tilde \ell_3 \tilde e_3 + \mathrm{h.c.} ~,
\end{equation}
with $A_t$, $A_b$, and $A_\tau$ also of order $\tilde m$. This assumption will be relaxed in section~\ref{sec:Aterms}.

In contrast, we assume that the scalar masses are \emph{completely anarchical}, with a texture
\begin{equation}
\label{eq:scalar-mass}
\tilde m^2_{q,u,d,\ell,e}\sim 
\begin{pmatrix} 
1 & 1 & 1 \cr
1 & 1 & 1 \cr
1 & 1 & 1 
\end{pmatrix}\,.
\end{equation}
As a result, after SUSY breaking the flavor symmetry is completely broken.  

\begin{figure}[t]
\centering
\includegraphics[width=0.3\textwidth]{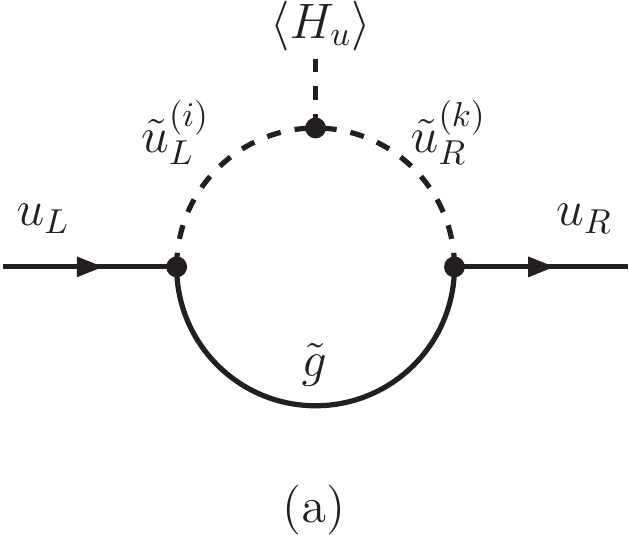} ~~~
\includegraphics[width=0.3\textwidth]{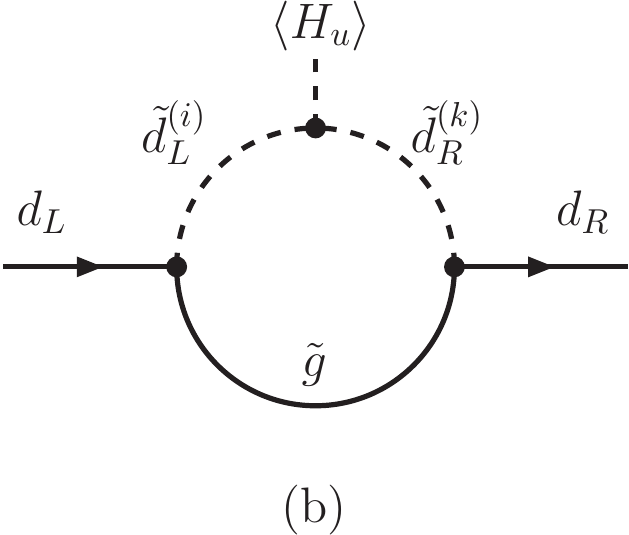} ~~~
\includegraphics[width=0.3\textwidth]{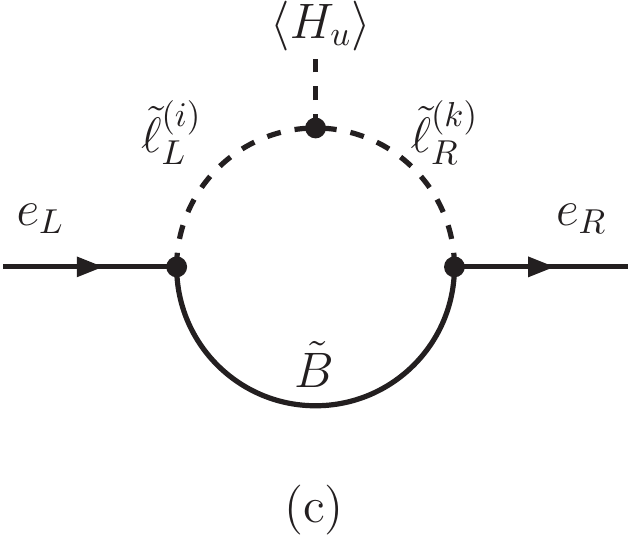}
\caption{Feynman diagrams that give the dominant 1-loop contributions to the first generation fermion masses. Shown are contributions to the (a) up quark mass, (b) down quark mass, and (c) electron mass. The indices $i, k = 1, 2 ,3$~ run over the mass eigenstates of left- and right-handed sfermions. Contributions to second generation masses as well as flavor off-diagonal masses are generated by analogous diagrams.}
\label{fig:diagrams1}
\end{figure}

Flavor violation will be mediated to the first and second generation fermions by loops of squarks and sleptons and the full Yukawa matrix is generated at 1-loop. The dominant diagrams are shown in figure~\ref{fig:diagrams1}. We note that in the case mixing in the up-top and charm-top sector, quark field renormalization diagrams contribute at the same order as the vertex correction shown in figure~\ref{fig:diagrams1}(a). However, the affect of these 1-3 and 2-3 entries on the fermion mass spectrum will be suppressed by the large top mass and will not affect the discussion of section~\ref{sec:mechanism2}.
In the down and lepton sectors, the field renormalization diagrams are suppressed by $1/\tan\beta$ compared to the vertex corrections and can be neglected.
We will now examine the fermion mass matrices that result from this minimal setup in greater detail and show that the mass spectrum is fully hierarchical.

\subsection{A Mechanism for Hierarchy from Anarchy}\label{sec:mechanism2}

Given our setup above it is clear that there will be a hierarchy of at least a loop factor between the third generation and the lighter two. One may worry, however, that because both the first and second generation masses are generated at 1-loop, they will be roughly degenerate and that additional flavor structure in equation~(\ref{eq:scalar-mass}) will be needed to explain the first-second generation mass splittings. We show here that this is not the case. In fact, even with fully anarchical scalar masses, \emph{our model is incapable of producing a degenerate electron and muon, and a mass splitting of order 100 is typical}. The same holds for quarks. 


For concreteness, let us consider the mass matrix of up-type quarks. The dominant contribution to the quark mass matrix, $m_u$, has the form
\begin{equation}
m_u^{ij} =m_\mathrm{tree}^{ij} + \Delta m_\mathrm{loop}^{ij}\,.
\end{equation}
By construction, $m_\mathrm{tree}$ is a matrix of rank 1 which can be written in a basis in which only its bottom-right entry is non-zero as in equation~(\ref{eq:yukawa-tree}). We will now show that $\Delta m_\mathrm{loop}$ is approximately a rank 1 matrix that is not aligned with $m_\mathrm{tree}$ and thus gives a mass dominantly to the second generation. The first generation mass arises only to the degree that~$\Delta m_\mathrm{loop}$ deviates from rank 1.

Evaluating the loop integral from diagram (a) in figure~\ref{fig:diagrams1}, we can write the loop induced mass matrix~$\Delta m_\mathrm{loop}$ as 
\begin{eqnarray}
\frac{\Delta m^{ij}_\mathrm{loop}}{v_u}& \propto&
Y_t (A_t+ \mu/\tan\beta) \times
\int \!\! \frac{d^4 k}{(2\pi)^4}
\sum_{mn}
\frac{ Z^L_{3n}Z^{L\dagger }_{ni}}{(k^2+\tilde m_{Ln}^2)}
\frac{ Z^R_{3m}Z^{R\dagger }_{mj}}{(k^2+\tilde m_{Rm}^2)}
\frac{ m_{\tilde g} }{(k^2+m_{\tilde g}^2)} ~,
\end{eqnarray}
where the $Z^{L,R}_{in}$'s are the mixing matrices from the left-handed and right-handed squarks,
and we have already performed a Wick rotation.
It is useful to re-write the loop induced mass as follows
\begin{equation}
\label{eq:int}
\frac{\Delta m^{ij}_\mathrm{loop}}{v_u} \propto
\int_0^\infty \!\! d k \, k^3\,
f_L^i(k)  f_R^j(k)
\frac{m_{\tilde g}}{(k^2+m_{\tilde g}^2)}  ~,
\end{equation}
where
\begin{equation}
\label{eq:f}
f_A^i(k) \equiv \sum_n
\frac{  Z^A_{3n}Z^{A\dagger }_{ni}}{ (k^2+\tilde m_{An}^2)} ~,
\end{equation}
and $A=L,R$. In equation~(\ref{eq:int}) we also converted the $d^4k$ integral to an integral over $dk$ (recall that we already Wick rotated). The object $f_A^i$ is a vector in flavor space which depends on $k^2$ and decides which fermion receives a mass from any given loop scale. 
It is interesting to note that for large loop momenta, $k^2\gg\tilde m^2$, $f^i_A$ points purely in the third generation direction due to the unitarity of the mixing matrix $Z_{in}$. To further investigate the properties of $\Delta m^{ij}_\mathrm{loop}$, we normalize the $f^i_A$'s to unit vectors and rewrite
\begin{equation}
\label{eq:fermion-mass}
\frac{\Delta m^{ij}_\mathrm{loop}}{v_u} \propto
\int_0^\infty \!\! dk\, 
|\Delta m(k)| \,
\hat f_L^i(k)\,  \hat f_R^j(k) ~,
\end{equation}
where the unit vectors are
 \begin{equation}
 \hat f_A^i\equiv \frac{f^i_A}{|f_A|} \qquad \mbox{with} \qquad 
  |f_A|^2\equiv \sum_i |f_A^i|^2 = \sum_n\frac{Z^A_{3n} Z^{A\dagger}_{n3}}{(k^2+m_{An}^2)^2}\,.
  \end{equation}
In the last step we have used the unitarity of the squark mixing matrix. The magnitude of the mass contributed by the momentum shell $k^2$ is
\begin{equation}
\label{eq:deltamk}
|\Delta m(k)| \equiv \frac{k^3|f_L||f_R|}{k^2+\tilde m_g^2}\,,
\end{equation}
which is invariant under flavor symmetries. 
There are two key points to notice about equations~(\ref{eq:fermion-mass}) and~(\ref{eq:deltamk}):
\begin{itemize}
\item The loop \emph{integrand} has the form $\hat f^i_L \hat f^j_R$, which is a rank 1 matrix, and thus respects a $U(2)_L\times U(2)_R$ symmetry. Because the vectors $\hat f_A$ change with $k$, the symmetry is broken ``collectively'' by the integration over all momentum scales . In other words, every fixed slice of loop momentum contributes mass to just one linear combination of fermions, and the integral is required to give mass to other fermions. 
\item  The magnitude of the loop integrand, $|\Delta m(k)|$ is a peaked function, dominated at the SUSY breaking scale $\tilde m$. It rises like $k^3$ for low $k$ and falls like $k^{-6}$ at high $k$'s. This means that the radiatively generated fermion mass will be approximately aligned with the integrand evaluated at the scale $k^2\sim \tilde m^2$.
\end{itemize}
Combining these observations gives us the following narrative for the hierarchy among the second and first generations: the loop momenta that dominate the loop integral, which are of order $k^2\sim \tilde m^2$, give a mass to  just one linear combination of fermions. We can define this combination as the second generation.  
As we vary $k$ away from $\tilde m$, by the time the $f^i_A$'s are sufficiently mis-aligned to contribute mass to the first generation the loop integrand is already small and the first generation mass is suppressed\footnote{The underlying mechanism is reminiscent of ``twisted split fermions''~\cite{Grossman:2004rm} in which fermion mass hierarchies are generated from localization in an extra dimension despite the fact that no flavor symmetry is preserved in the bulk of the extra dimension.}. Note that this mechanism is independent of the flavor structure in the scalar sector and is present even when the scalar masses are fully anarchic. 
It is worth noting that the generated hierarchy does not arise from a small parameter in our theory, but from the properties of  the loop integral and collective symmetry breaking in loop momentum space. 

\begin{figure}[t]
\centering
\includegraphics[width=0.54\textwidth]{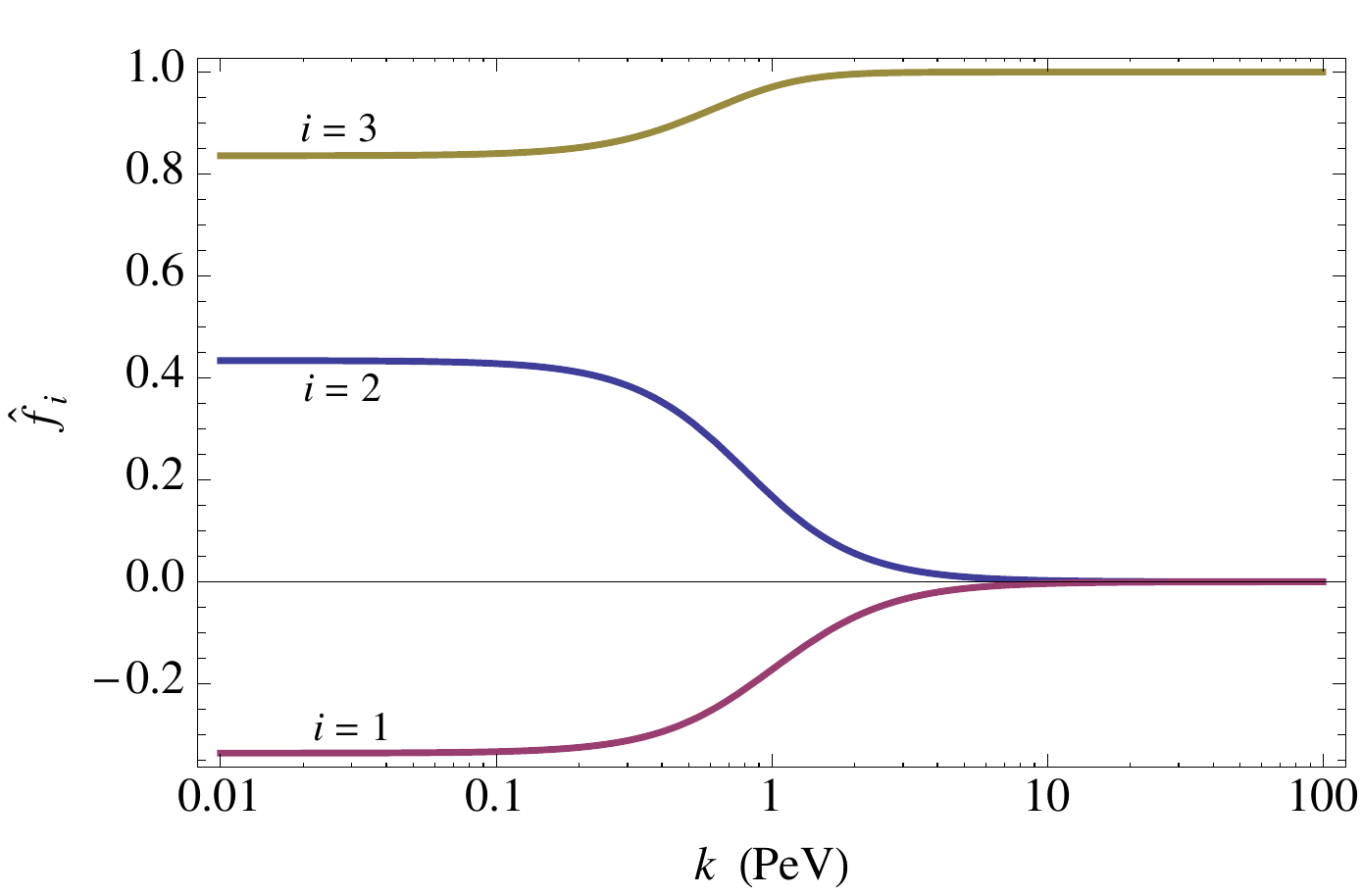}\\
\hspace{0.1cm} \includegraphics[width=0.52\textwidth]{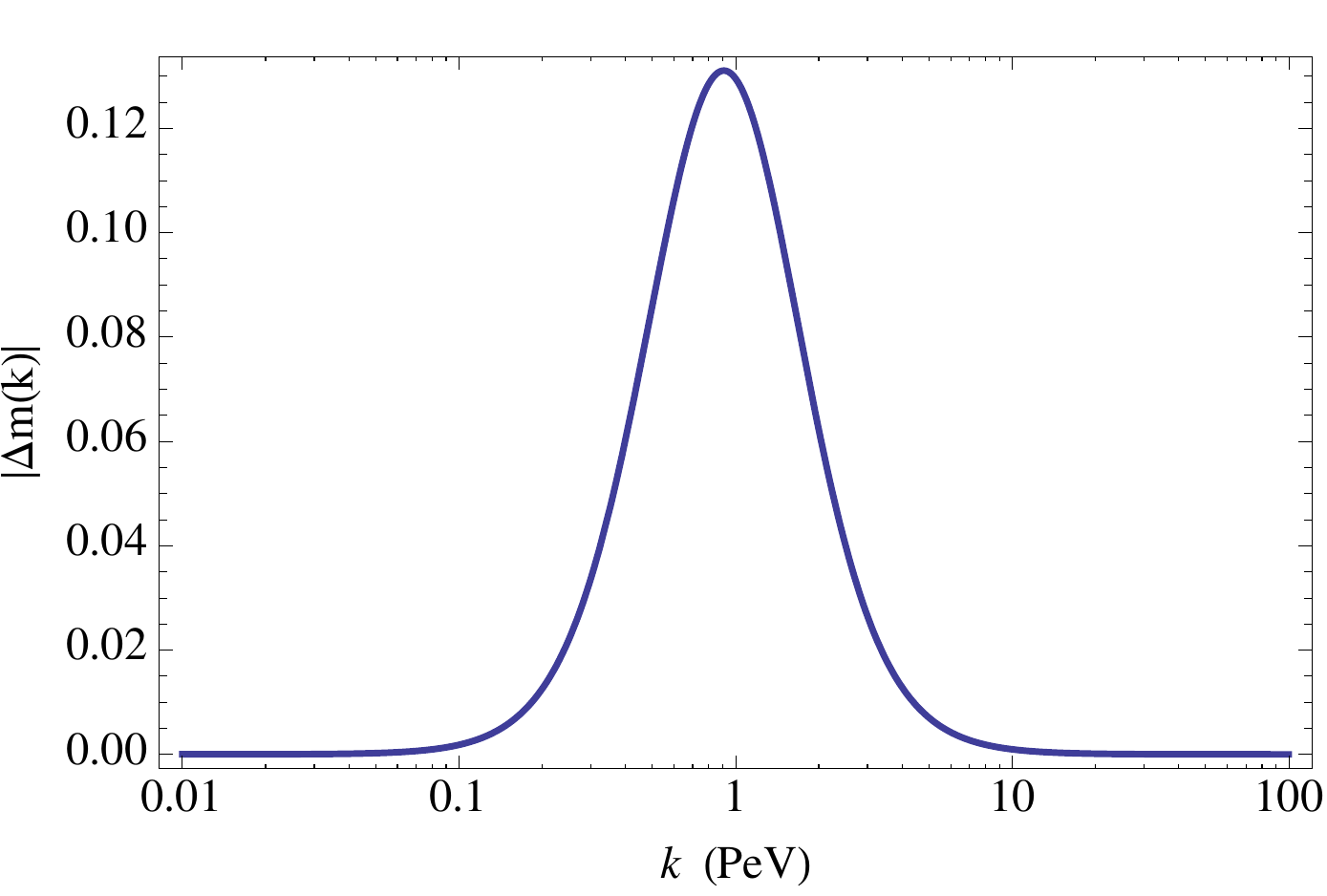}\\
\caption{The important quantities in setting the mass hierarchy in our mechanism for a generic choice of an anarchic scalar mass matrix. At every loop momentum $k$ only one linear combination of fermions receives mass. The flavor vector $\hat f^i$ determines the combination of fermions  while the function $|\Delta m_k|$ sets the magnitude of the mass. Due to the strong localization of $|\Delta m_k|$ the resulting mass matrix is approximately rank 1.}
\label{fig:fi}
\end{figure}

To demonstrate this mechanism more visually, we plot $\hat f^i$ and $|\Delta m(k)|$ for a generic
choice\footnote{The particular mass matrix we chose here is $\tilde m_{11}^2=.8$, $m_{22}^2=1$, $m_{33}^2=1.2$, $m_{12}^2=.3$, $m_{23}^2=.2$, $m_{13}^2=-.4$, in units of PeV$^2$. The gluino mass is chosen to be 1~PeV. The qualitative features discussed here do not depend on this choice.}   of scalar masses in figure~\ref{fig:fi} (where we assumed $\tilde m^2_L=\tilde m^2_R$ for simplicity).  We see that the contribution to the fermion mass is indeed ``localized'' to momenta of order $k \sim \tilde m$. The unit vectors $\hat f^i$ do not change drastically in this region of loop momentum and define the second generation. The first generation receives a mass only to the degree that the $f^i$'s manage to change in the 1-2 plane in this region. 

Having a qualitative understanding of our mechanism for fermion hierarchy, we continue to investigate the spectrum more quantitatively in the next section.

\section{The Fermion Spectrum in the Minimal Model}\label{sec:spectrum}

We would like to get a clearer understanding of the fermion spectrum in our model and to further quantify the suppression effect of first generation masses. In subsection~\ref{sec:minimal_analytic} we will do this first analytically. To keep the formulae manageable we will work in the mass insertion approximation and take into account only the dominant 1-loop contributions shown in figure~\ref{fig:diagrams1}. 
In subsection~\ref{sec:stability} we will consider the constraints that vacuum stability places on the parameter space. In subsection~\ref{sec:minimal_numeric} we will perform a numerical scan in the mass basis to see that the mass insertion results are robust.

\subsection{Spectrum in the Minimal Model: Analytic Results} \label{sec:minimal_analytic}

We parameterize the soft masses of squarks and sleptons as 
\begin{subequations}
\begin{equation}
 m_Q^2 = m_{\tilde q}^2 ( 1\!\!1 + \delta_q^L)~,~~ m_U^2 = m_{\tilde u}^2 ( 1\!\!1 + \delta_u^R)~,~~ m_D^2 = m_{\tilde d}^2 ( 1\!\!1 + \delta_d^R)~,
\end{equation}
\begin{equation}
 m_L^2 = m_{\tilde \ell}^2 ( 1\!\!1 + \delta_\ell^L)~,~~ m_E^2 = m_{\tilde e}^2 ( 1\!\!1 + \delta_\ell^R)~,
\end{equation}
\end{subequations}
where $\delta_f^A$ are dimensionless matrices that encode the flavor breaking and mass splittings, and whose elements are all allowed to be $O(1)$. 
Following~\cite{ArkaniHamed:1996zw}, we use $U(2)^5$ flavor rotations that leave the Yukawa couplings invariant, and rotate to a basis where the $(1,3)$ and $(3,1)$ entries of all $\delta$'s vanish
\begin{subequations}
\begin{equation}
 \delta_q^L = \begin{pmatrix} \delta_{11}^L & \delta_{12}^L & 0 \\ \delta_{21}^L & \delta_{22}^L & \delta_{23}^L \\ 0 & \delta_{32}^L & \delta_{33}^L \end{pmatrix} ~,~~ \delta_u^R = \begin{pmatrix} \delta_{uu}^R & \delta_{uc}^R & 0 \\ \delta_{cu}^R & \delta_{cc}^R & \delta_{ct}^R \\ 0 & \delta_{tc}^R & \delta_{tt}^R \end{pmatrix} ~,~~ \delta_d^R = \begin{pmatrix} \delta_{dd}^R & \delta_{ds}^R & 0 \\ \delta_{sd}^R & \delta_{ss}^R & \delta_{sb}^R \\ 0 & \delta_{bs}^R & \delta_{bb}^R \end{pmatrix} ~,
\end{equation}
\begin{equation}
 \delta_\ell^L = \begin{pmatrix} \delta_{ee}^L & \delta_{e\mu}^L & 0 \\ \delta_{\mu e}^L & \delta_{\mu\mu}^L & \delta_{\mu\tau}^L \\ 0 & \delta_{\tau\mu}^L & \delta_{\tau\tau}^L \end{pmatrix} ~,~~ \delta_\ell^R = \begin{pmatrix} \delta_{ee}^R & \delta_{e\mu}^R & 0 \\ \delta_{\mu e}^R & \delta_{\mu\mu}^R & \delta_{\mu\tau}^R \\ 0 & \delta_{\tau\mu}^R & \delta_{\tau\tau}^R \end{pmatrix} ~.
\end{equation}
\end{subequations}
Assuming for simplicity a SUSY spectrum with degenerate sfermions and gauginos, $m_{\tilde q} = m_{\tilde u} = m_{\tilde d} = m_{\tilde \ell} = m_{\tilde e} = m_{\tilde B} = m_{\tilde W} = m_{\tilde g}$, we find the following fermion mass spectrum at leading order in the mass insertion approximation
\begin{subequations}
\begin{eqnarray} \label{eq:radmass1}
 \frac{m_c}{m_t} &\simeq& \frac{\alpha_s}{4\pi} \frac{2}{9} \frac{|m_{\tilde g} A_t|}{m_{\tilde q}^2}  |\delta_{23}^L\delta_{tc}^R|~, \\ \label{eq:radmass2}
 \frac{m_u}{m_t} &\simeq& \frac{\alpha_s}{4\pi} \frac{2}{225} \frac{|m_{\tilde g} A_t|}{m_{\tilde q}^2} |\delta_{12}^L\delta_{cu}^R\delta_{23}^L\delta_{tc}^R|~, \\ \label{eq:radmass3}
 \frac{m_s}{m_b} &\simeq& \frac{\alpha_s}{4\pi} \frac{2}{9} \frac{|m_{\tilde g} \mu| }{m_{\tilde q}^2} \frac{t_\beta}{|1 + \epsilon_b t_\beta|} |\delta_{23}^L\delta_{bs}^R| ~, \\  \label{eq:radmass4}
 \frac{m_d}{m_b} &\simeq& \frac{\alpha_s}{4\pi} \frac{2}{225} \frac{|m_{\tilde g} \mu| }{m_{\tilde q}^2} \frac{t_\beta}{|1 + \epsilon_b t_\beta|} |\delta_{12}^L\delta_{sd}^R\delta_{23}^L\delta_{bs}^R| ~, \\ \label{eq:radmass5}
 \frac{m_\mu}{m_\tau} &\simeq& \frac{\alpha_1}{4\pi} \frac{1}{12} \frac{|m_{\tilde B} \mu|}{m_{\tilde \ell}^2} \frac{t_\beta}{|1 + \epsilon_\tau t_\beta|}  |\delta_{\mu\tau}^L\delta_{\tau\mu}^R| ~, \\  \label{eq:radmass6}
 \frac{m_e}{m_\tau} &\simeq& \frac{\alpha_1}{4\pi} \frac{1}{300} \frac{|m_{\tilde B} \mu|}{m_{\tilde \ell}^2} \frac{t_\beta}{|1 + \epsilon_\tau t_\beta|} |\delta_{e\mu}^L\delta_{\mu e}^R\delta_{\mu\tau}^L\delta_{\tau\mu}^R| ~.
\end{eqnarray}
\end{subequations}
The $\epsilon_b$ and $\epsilon_\tau$ factors in the down-type quark and lepton mass ratios arise from the loop corrections to the tree-level bottom and tau masses. As they are $\tan\beta$ enhanced, they can have a non-negligible effect\footnote{In fact, for sufficiently large $\tan\beta$ these loop corrections can dominate the bottom and tau masses~\cite{Dobrescu:2010mk,Altmannshofer:2010zt}. In this case the top-bottom and top-tau hierarchies are also explained by loops. This is indeed one of the interesting regions of parameter space of our model.}. Using the same approximations as above, we have
\begin{equation} \label{eq:eps}
 \epsilon_b \simeq \frac{\alpha_s}{4\pi} \frac{4}{3} \frac{m_{\tilde g} \mu}{m_{\tilde q}^2} ~,~~
 \epsilon_\tau \simeq \frac{\alpha_2}{4\pi} \frac{3}{4} \frac{m_{\tilde W} \mu}{m_{\tilde \ell}^2}  ~.
\end{equation}

\begin{figure}[t]
\centering
\includegraphics[width=0.6\textwidth]{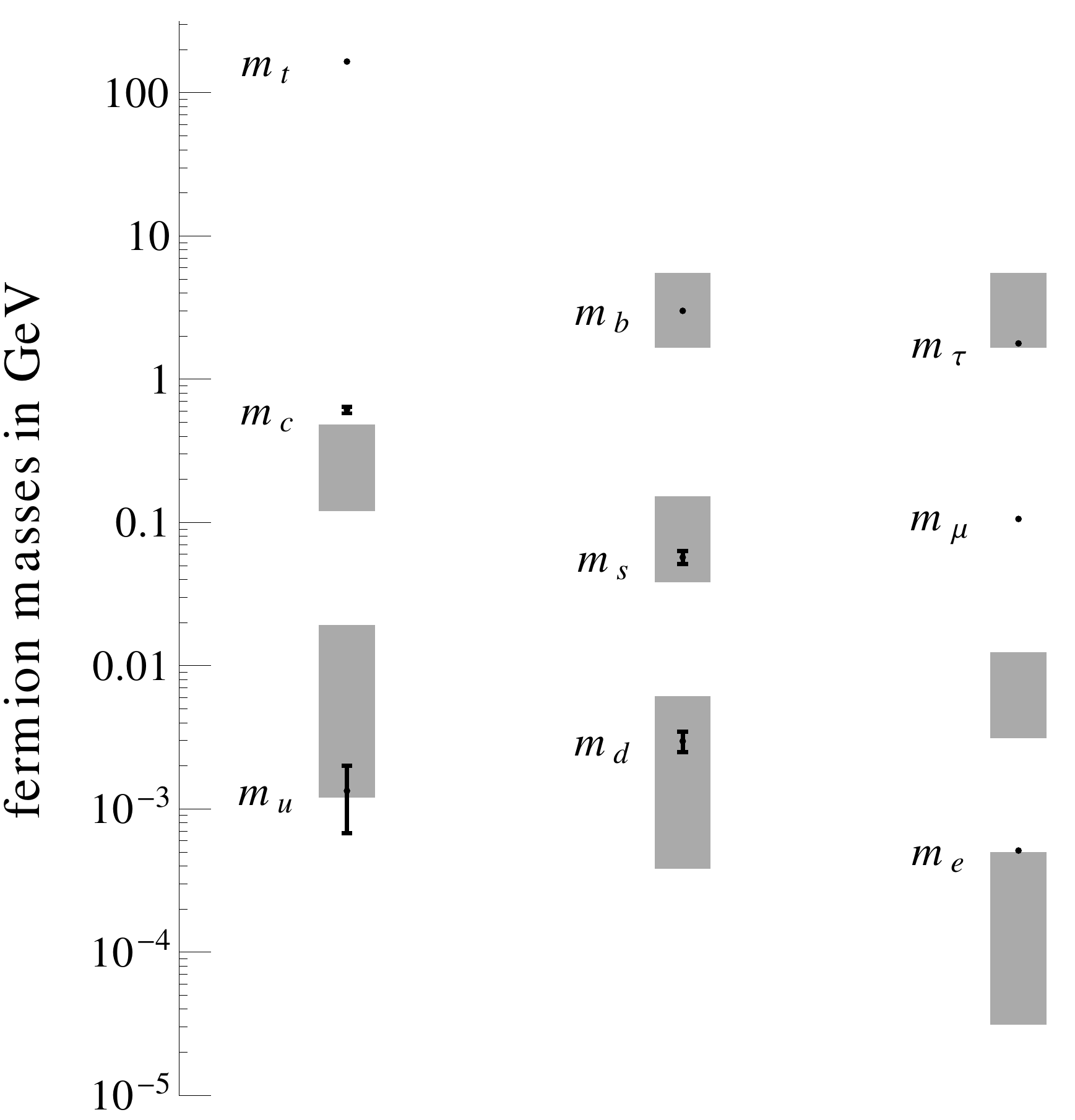}
\caption{Generic predictions for the hierarchies in the SM fermion masses in the minimal model. The gray regions are obtained from a leading order mass insertion approximation. All SUSY masses are set to a common value $\tilde m = 1$~PeV, the stop trilinear coupling is $A_t = 3 \tilde m$ and $\tan\beta$ is varied between $30 < \tan\beta < 100$. The mass insertions are varied in the range $0.5 < \delta^A_f < 1$. Shown are also the observed $\overline{\text{MS}}$ fermion masses at the electro-weak scale, with their $2\sigma$ uncertainties.}
\label{fig:chart_MSSM}
\end{figure}

The generic predictions of the loop induced spectrum~(\ref{eq:radmass1}) -~(\ref{eq:radmass6}) are summarized in figure~\ref{fig:chart_MSSM} and compared to the experimental values of the quark and lepton masses.
The gray regions are obtained by setting all SUSY masses to a common value $\tilde m = 1$~PeV and the stop trilinear coupling to $A_t = 3 \tilde m$. All mass insertions are varied between $0.5 < \delta^A_f < 1$ and $30 < \tan\beta < 100$.
We now discuss some of the features of this spectrum in more detail.
\paragraph*{\underline{Third generation masses:}}
The top Yukawa coupling is of order one by construction. The masses of the bottom and the tau are $1/\tan\beta$ suppressed. It is thus attractive to take a large value for $\tan\beta$, allowing for bottom and tau Yukawa couplings of order one. 
\paragraph*{\underline{Second-third generation mass splitting:}} 
The second generation mass, as expected, is a 1-loop factor below the third generation. It is, however, worth taking note of the $\tan\beta$ scaling. The loop induced second generation mass is proportional to the trilinear connecting the up-type Higgs to the corresponding scalars. For the up-type quarks this is the supersymmetry breaking $A_t$, while for the down-type quarks and leptons it is dominated by the non-holomorphic trilinear, scaling like $\mu\tan\beta$. We thus find that the $\tan\beta$ scaling which was suppressing the bottom and $\tau$ masses is un-done for the second generation. The strange quark and muon masses are thus parametrically similar to the charm quark mass (at least as far as the $\tan\beta$ scaling is concerned). We view this as a desirable feature.

The trilinear couplings discussed here are constrained by vacuum stability, and the naive hierarchy between third and second generation fermions are a bit too large. We will address this in detail later.

\paragraph*{\underline{First-second generation mass splitting:}}
As anticipated, a hierarchy between the first and second generation masses is generated automatically even for a completely anarchic flavor structure of the sfermions
\begin{equation} \label{eq:hierachy}
 \frac{m_u}{m_c} \simeq \frac{1}{25} ~ |\delta_{12}^L\delta_{cu}^R| ~,~~ 
 \frac{m_d}{m_s} \simeq \frac{1}{25} ~ |\delta_{12}^L\delta_{sd}^R| ~,~~
 \frac{m_e}{m_\mu} \simeq \frac{1}{25} ~ |\delta_{e\mu}^L\delta_{\mu e}^R| ~.
\end{equation}
Comparing to the experimentally determined ratios collected in the appendix~(\ref{eq:masses1})--(\ref{eq:masses3}), we see that adjusting the involved mass insertions by suitable $O(1)$ factors allows to accommodate the observed ratios of first and second generation masses easily. 

\subsection{Considerations of Vacuum Stability} \label{sec:stability}

Before plugging in numbers and getting detailed spectra, we pause to consider the constraint that vacuum stability places on our parameter space. A supersymmetric theory invariably introduces many colored and charged scalars and care must be taken to ensure our electro-weak breaking vacuum is stable, or at least meta-stable.

The first consideration is to ensure none of the charged and colored scalars are tachionic, namely, the eigenvalues of the anarchic scalar mass-squared matrices must all be positive. As off-diagonal entries in a hermitian matrix always cause a lower determinant, it is clear that requiring positive eigenvalues bounds them from above. The analytic form of the upper bound on any particular flavor violating mass insertion depends on the other entries of the mass matrix and is not enlightening for our discussion. In practice we find that the flavor violation mass insertions tend to be less than 1/2. In the next subsection we will take a pragmatic approach, and scan over all anarchic scalar mass matrices. Matrices that produce one or more tachyons in their spectrum are simply discarded.
We note that the 2-3 and 1-2 fermion mass hierarchies are set by the off-diagonal mass insertions $\delta_{23}$ and $\delta_{12}$ respectively. The vacuum stability requirement thus pushes our fermion mass spectrum to be somewhat more hierarchical than it would have been without this constraint.

Even if all scalars have a positive mass-squared, we also need to consider more distant color and charge breaking vacua to which our vacuum can decay. The radiatively induced fermion masses depend sensitively on the stop trilinear coupling $A_t$ (in the case of the up-type quarks) and on the non-holomorphic trilinear which is set by $\mu$ (in the case of down-type quarks and the leptons). 
Like any trilinear term, these interactions may induce new minima, or runaway directions.
In particular, it is well known~\cite{Claudson:1983et,Casas:1995pd}, that a large stop trilinear can induce a deep charge and color breaking minimum in the MSSM scalar potential along the D-flat direction $\tilde t_L = \tilde t_R = H_u$. Requiring that the electro-weak vacuum is the deepest minimum in the MSSM scalar potential leads to a bound on the stop trilinear coupling
\begin{equation} \label{eq:Atbound}
 |A_t|^2 \lesssim 3 (m_{\tilde t_L}^2 + m_{\tilde t_R}^2 + m_{H_u}^2 + |\mu|^2) \simeq 3 (m_{\tilde t_L}^2 + m_{\tilde t_R}^2)~.
\end{equation}
This bound is derived under the assumption of flavor conserving squark soft masses. In the general case with $O(1)$ squark mixing, the bound can be stronger as discussed in the appendix~\ref{app:vac}. If one allows the electro-weak vacuum to decay, and only demands that its life-time be longer than the age of the universe, the bound gets slightly relaxed~\cite{Kusenko:1996jn,Blinov:2013fta}. The upper bound on the stop trilinear coupling constrains the size of the radiatively induced charm and up masses. Note that the constraint~(\ref{eq:Atbound}) on the relative size of the trilinear coupling with respect to the soft masses is independent on the overall mass scale. This statement remains approximately true also if one allows for a meta-stable vacuum.

For large values of~$\mu$, vacua with sbottom or stau vevs can appear in the MSSM scalar potential~\cite{Rattazzi:1996fb,Hisano:2010re,Altmannshofer:2012ks,Carena:2012mw}. Even if there are no exact D-flat directions involving only the up-type Higgs and sbottoms or staus, the trilinear couplings of the up-type Higgs with the sbottoms and staus
can lead to charge and color breaking minima, if the 
product of Higgsino mass and bottom/tau Yukawa is of the order of the squark/slepton soft masses. Simple analytical bounds for the Higgsino mass equivalent to~(\ref{eq:Atbound}) can only be obtained for a fixed direction in field space. For any given field direction, the bounds are only necessary but usually not sufficient to guarantee the absence of charge and color breaking minima. In the large $\tan\beta$ regime, we find that in the optimized field direction $H_u$ is usually a bit larger than the sfermion fields. For definiteness we chose here exemplarily $\tilde \tau_L = \tilde \tau_R = H_u/\sqrt{2}$ and $\tilde b_L = \tilde b_R = H_u/\sqrt{2}$. In this case we find the necessary conditions
\begin{subequations}
\begin{eqnarray} \label{eq:mubound}
 |\mu|^2 &\lesssim& (m_{\tilde \tau_L}^2 + m_{\tilde \tau_R}^2) \left[ \frac{1}{2} + \frac{9}{32} \frac{v^2}{m_\tau^2} \left|\frac{1}{t_\beta} + \epsilon_\tau \right|^2(g_1^2 + g_2^2) \right]~, \\ \label{eq:mubound2}
 |\mu|^2 &\lesssim& (m_{\tilde b_L}^2 + m_{\tilde b_R}^2) \left[ \frac{1}{2} + \frac{9}{32} \frac{v^2}{m_b^2} \left|\frac{1}{t_\beta} + \epsilon_b \right|^2(g_1^2 + g_2^2) \right]~,
\end{eqnarray}
\end{subequations} 
where we expressed the Yukawa couplings in terms of fermion masses, and included the $\tan\beta$ enhanced threshold corrections $\epsilon_\tau$ and $\epsilon_b$. Depending on the phase of the $\epsilon_i$, the threshold corrections can significantly strengthen or loosen the constraints. The fermion masses as well as the gauge couplings in~(\ref{eq:mubound}) and~(\ref{eq:mubound2}) should be evaluated at the PeV scale.
In the presence of generic flavor mixing of the squarks and sleptons, the bounds can become stronger as discussed in the appendix~\ref{app:vac}. Analogously to the case of the stop trilinear coupling, requiring only meta stability will relax the bounds slightly. Nonetheless, it is clear from~(\ref{eq:mubound}) and~(\ref{eq:mubound2}) that for large $\tan\beta \sim 50$ the Higgsino mass, $\mu$, cannot exceed the sbottom and stau masses by more than a factor of few.

How do these constraints affect the fermion mass spectrum?
To obtain a charm mass~(\ref{eq:radmass1}) in agreement with the data (see appendix~\ref{app:masses}), maximal mass insertions and a large $A_t$ term close to the vacuum stability bound~(\ref{eq:Atbound}) are required. A large enough strange quark mass~(\ref{eq:radmass3}) is only possible in the large $\tan\beta \sim 50$ regime and for a Higgsino mass at least of the order of the squark masses. Even for large $\tan\beta$ and maximal mass insertions in the slepton sector, the muon mass is typically one order of magnitude below the measured value. In order to obtain a muon mass from~(\ref{eq:radmass5}) in agreement with observation, the Higgsino mass needs to be
\begin{equation} \label{eq:mu}
 |\mu| \simeq \frac{m_\mu}{m_\tau} \frac{m_{\tilde \ell}^2}{|m_{\tilde B}|} \frac{4\pi}{\alpha_1} 12 \left|\frac{1}{t_\beta} + \epsilon_\tau \right| \frac{1}{|\delta_{\mu\tau}^L\delta_{\tau\mu}^R|} \sim 16 \times m_{\tilde \ell} ~,
\end{equation}
where in the last step we chose $\tan\beta = 50$, neglected the $\epsilon_\tau$ term and set all mass insertions generously to 1. 
A Higgsino mass as large as in~(\ref{eq:mu}) is strongly excluded by the vacuum stability requirements discussed above. One may hope for an accidental cancellation between the $1/t_\beta$ and the $\epsilon_\tau$ terms, which can bring equations~(\ref{eq:mubound}) and~(\ref{eq:mu}) into apparent agreement. We note first that such an accidental cancellation would be rare indeed, particularly in light of the fact that the phase of $\epsilon_\tau$ is arbitrary. In addition, we note that equation~(\ref{eq:mubound}) considers a particular direction in field space. This is not the direction of greatest instability in the case of extreme cancellations in the potential. In section~\ref{sec:mod} we will return to the problem of the muon mass in our model and propose solutions.

\subsection{Spectrum in the Minimal Model: Numerical Results} \label{sec:minimal_numeric}

The conclusions of subsections~\ref{sec:minimal_analytic} and~\ref{sec:stability} are based on the mass insertion approximation and include only the leading diagrams which contribute to fermion masses. One would like to explore the parameter space beyond these approximations and to investigate for example the possibility of cancellations among various diagrams.
To investigate our parameter space fully, in this subsection we will analyze the radiatively induced fermion masses numerically. We will do this without making use of the mass insertion approximation and taking into account the complete set of 1-loop contributions that consist of the gluino and bino loops shown in figure~\ref{fig:diagrams1}, as well as Wino loops, Higgsino loops, Wino-Higgsino loops, and Bino-Higgsino loops. 

\begin{figure}[t]
\centering
\includegraphics[width=\textwidth]{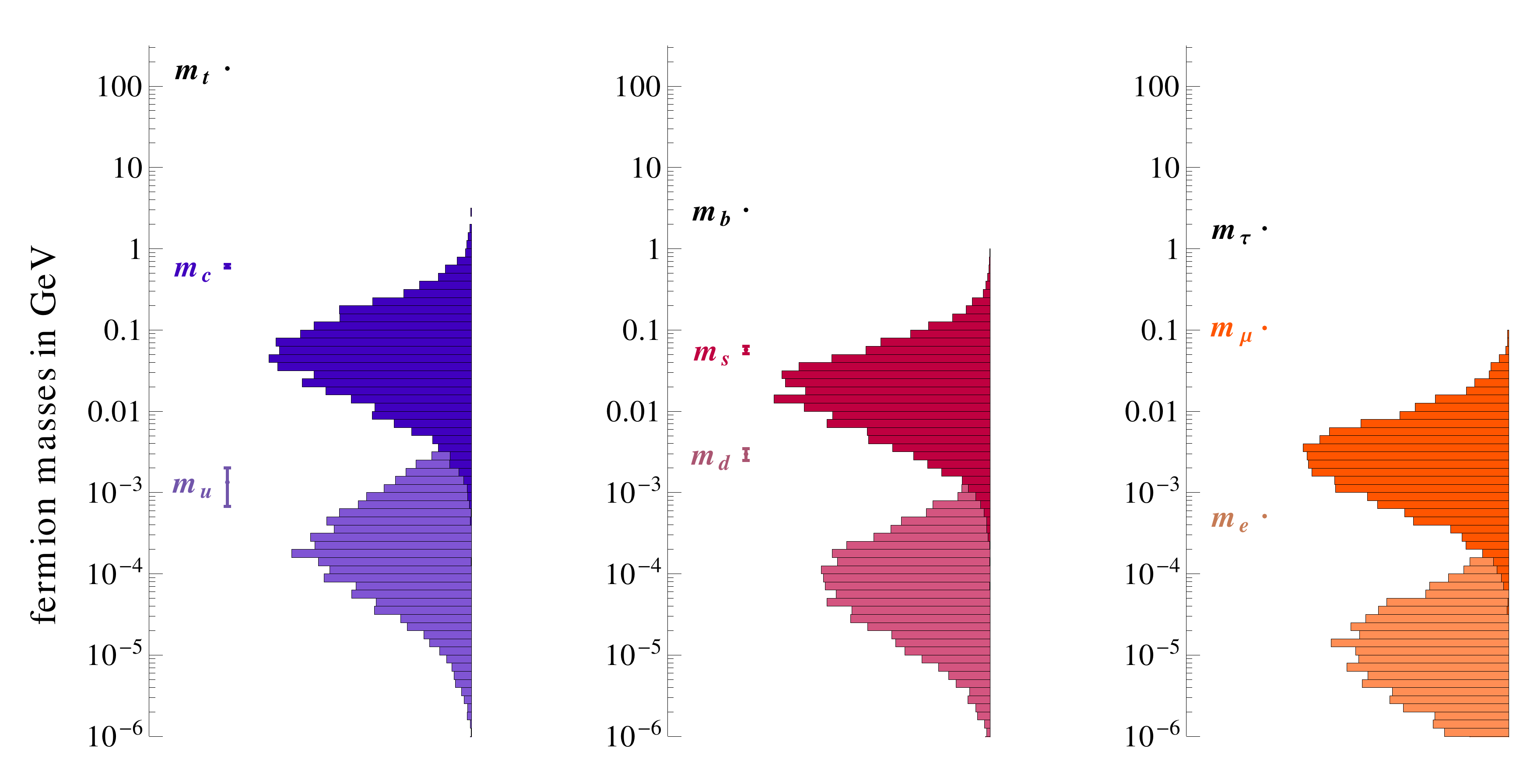}
\caption{Distributions of the fermion masses as obtained by a parameter scan of the minimal model (see text for details). The masses correspond to $\overline{\text{MS}}$ masses at the electro-weak scale. Shown are also the observed masses with $2\sigma$ uncertainties.}
\label{fig:scan_masses_MSSM}
\end{figure}

\begin{figure}[t]
\centering
\includegraphics[width=0.45\textwidth]{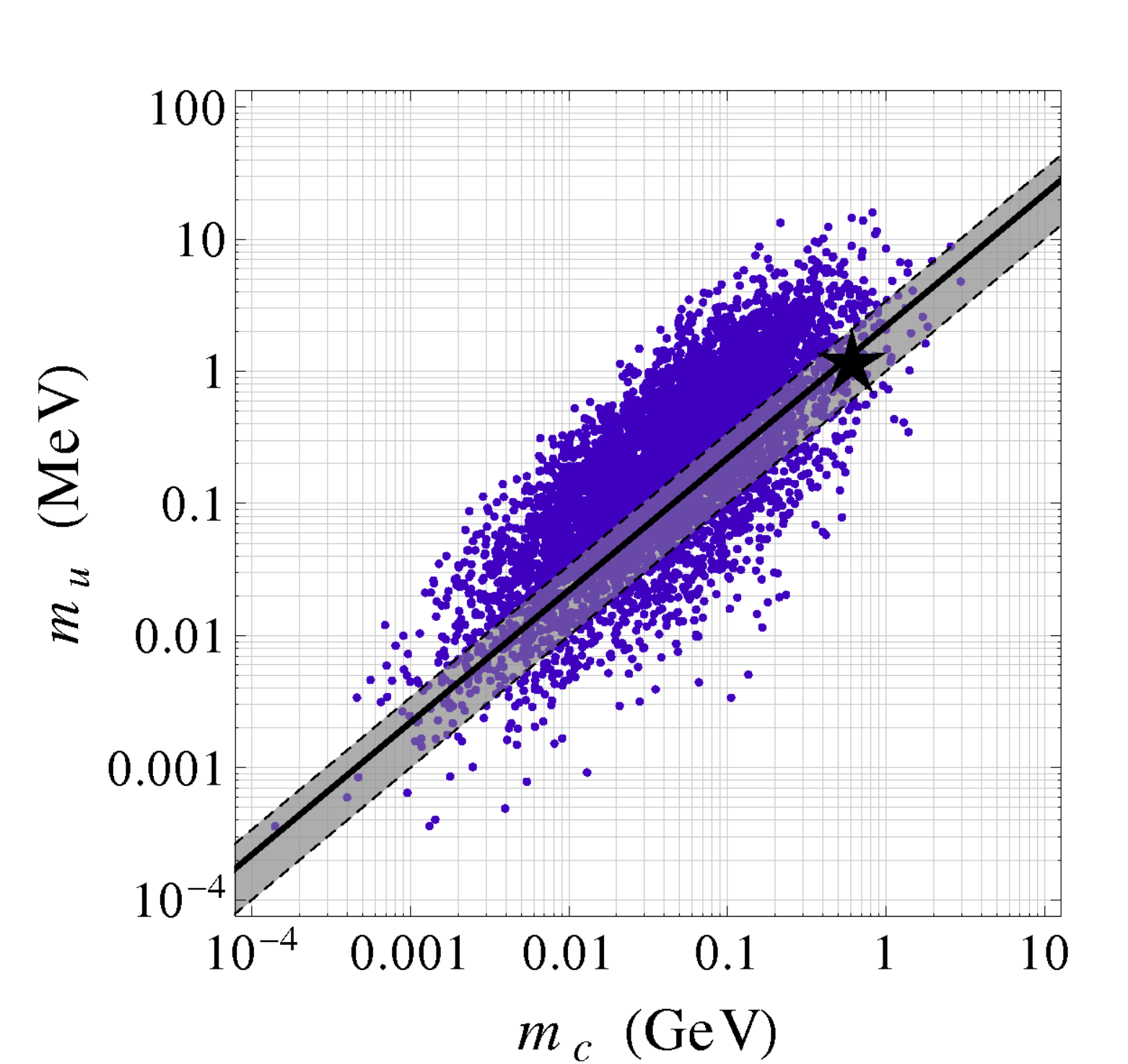} ~~~~~
\includegraphics[width=0.45\textwidth]{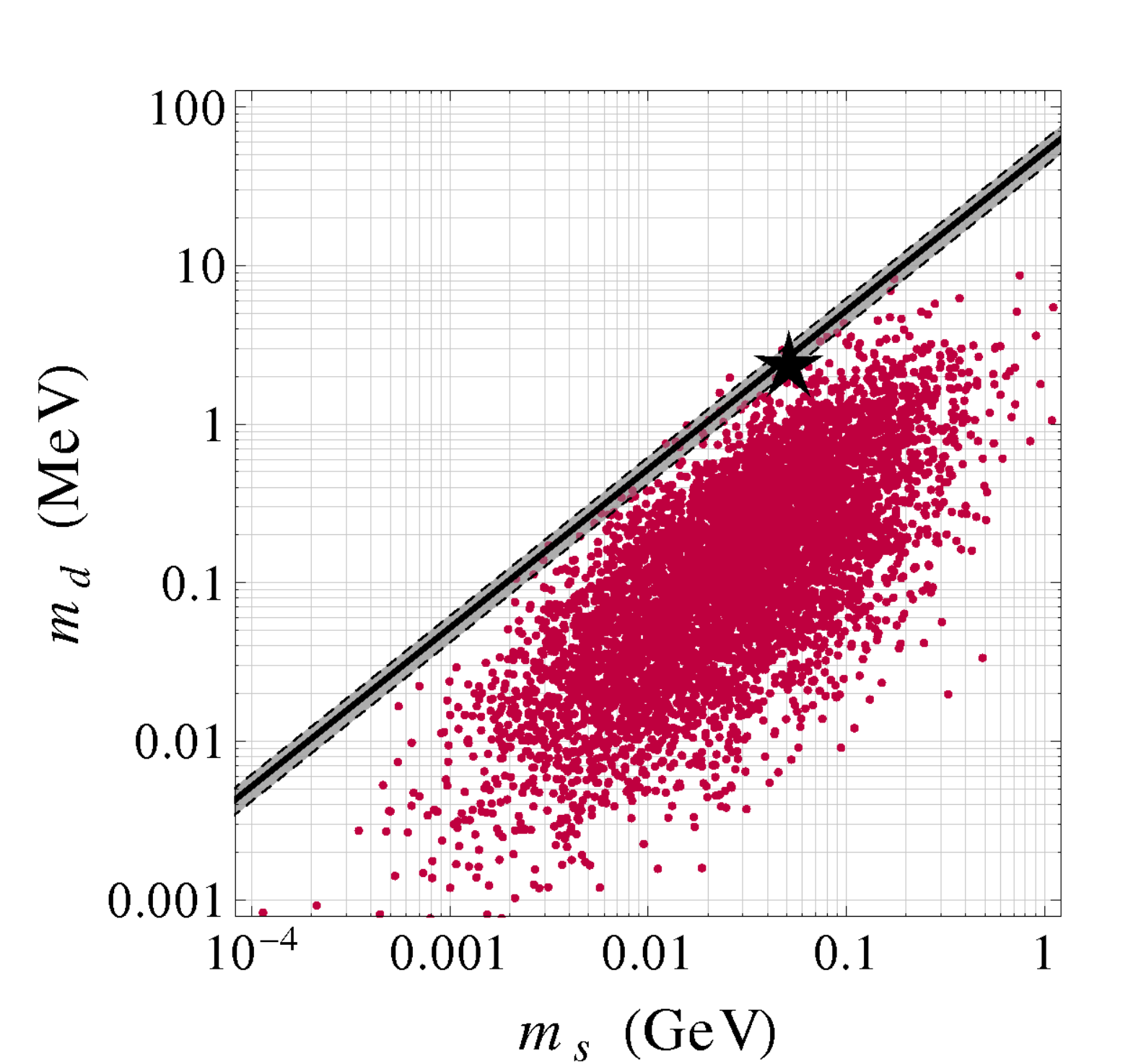} \\
\includegraphics[width=0.45\textwidth]{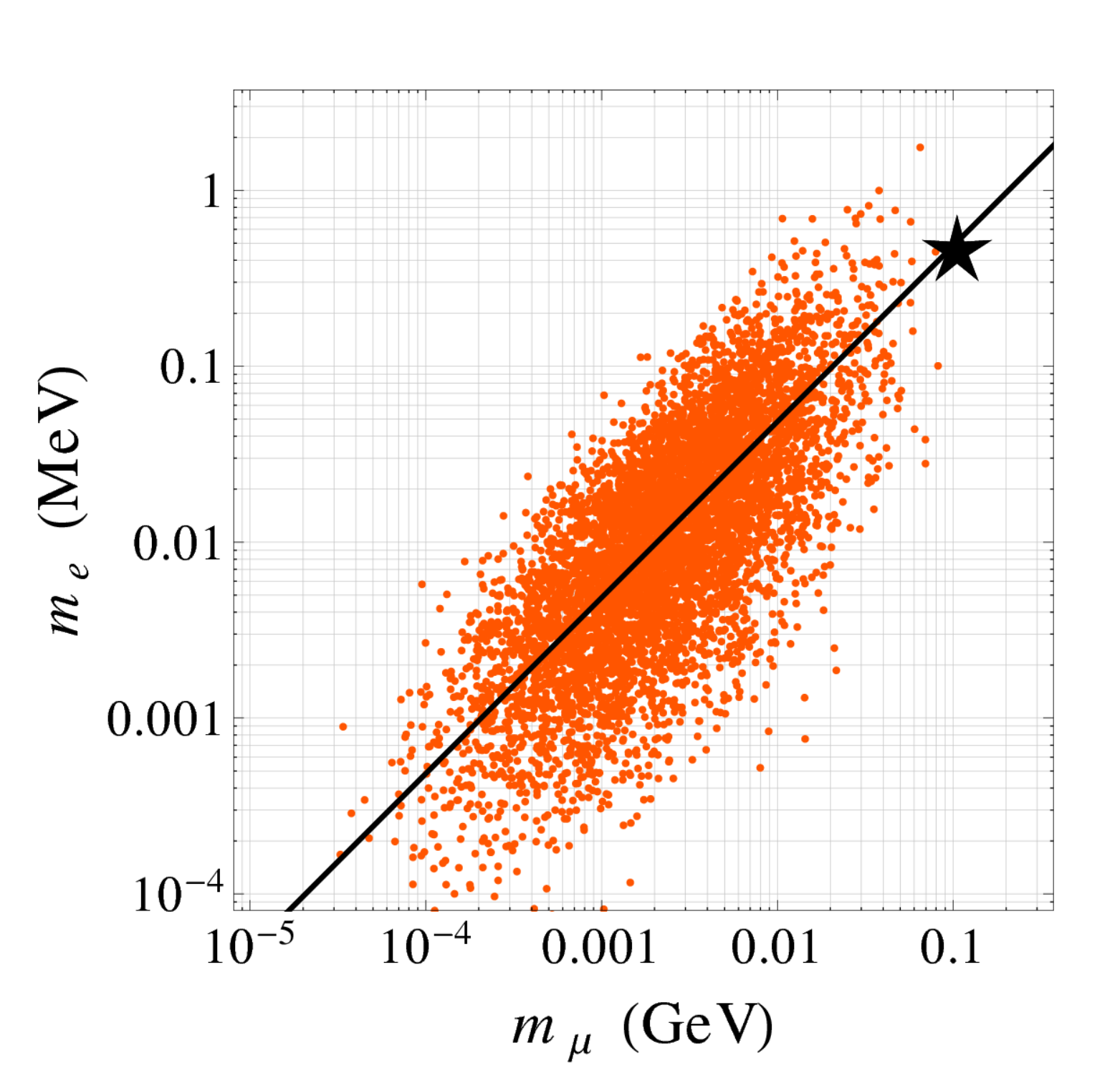}
\caption{Masses of the first and second fermion generations as obtained by a parameter scan of the minimal model (see text for details). The masses correspond to $\overline{\text{MS}}$ masses at the electro-weak scale. The black stars indicate the observed values. The diagonal bands show the observed ratios of masses with $2\sigma$ uncertainties.}
\label{fig:1st2nd_MSSM}
\end{figure}

In order to find the ranges of fermion masses that can be accommodated in the discussed setup, we perform a numerical scan of the model parameters, the results of which are shown in figures~\ref{fig:scan_masses_MSSM} and~\ref{fig:1st2nd_MSSM}.
Apart from small renormalization group effects to be discussed below, the loop induced fermion mass matrices are independent of the overall scale of SUSY particles $\tilde m$ that therefore can be set to an arbitrary value $\tilde m = 1$~PeV. We scan all mass parameters of the model (gaugino masses, Higgsino mass, trilinear couplings, diagonal and off-diagonal sfermion masses) within one order of magnitude around the common scale $[0.3 \tilde m, 3 \tilde m]$ using flat distributions. We also assign random phases between $0$ and $2\pi$ to all complex parameters. The third generation Yukawa couplings are fixed such that the third generation masses agree with the observed values. We impose the absence of tachyons as well as the approximate vacuum stability bounds in~(\ref{eq:Atbound}),~(\ref{eq:mubound}), and~(\ref{eq:mubound2}).

In comparing our results to observation we also take into account renormalization group effects. In particular, we evaluate the gauge couplings at the scale $\tilde m = 1$~PeV
\begin{equation}
 g_1(1 \text{PeV}) \simeq 0.38 ~,~~ g_2(1 \text{PeV}) \simeq 0.61 ~,~~ g_3(1 \text{PeV}) \simeq 0.81 ~.
\end{equation}
These values are obtained using SM 2-loop renormalization group equations (RGEs)~\cite{Machacek:1983fi,Machacek:1983tz,Machacek:1984zw,Luo:2002ey}.
Integrating out the SUSY particles, we obtain ratios of quark and lepton masses at the SUSY scale $\tilde m \simeq m_{\tilde q}, m_{\tilde\ell}$. In order to compare the ratios to the experimentally measured mass ratios given in the appendix~\ref{app:masses}, we run them down to the electro-weak scale using again SM 2-loop RGEs. At 1-loop, the effect of the gauge interactions cancel in the ratios. Only the top Yukawa leads to a correction of the ratios $m_u/m_t$ and $m_c/m_t$. In a 1-loop leading log approximation we find
\begin{equation} \label{eq:running}
 \frac{m_{u,c}(m_t)}{m_t(m_t)} \simeq \frac{m_{u,c}(\tilde m)}{m_t(\tilde m)} \left[ 1 + \frac{Y_t^2}{16\pi^2} \frac{3}{4} \log\left(\frac{\tilde m^2}{m_t^2}\right)  \right] ~.
\end{equation}
For $\tilde m = 1$~PeV, the RGE correction in~(\ref{eq:running}) is approximately $5\%$ and therefore hardly relevant. All other mass ratios are to an excellent approximation RGE invariant. 

The value of $\tan\beta$ is fixed to $\tan\beta = 60$ in the scan. Smaller values of $\tan\beta$ lead to smaller masses of the down-type quarks and leptons. Larger values of $\tan\beta$ lead to stronger constraints from stau and sbottom vevs. We find that $\tan\beta > 60$ does not lead to appreciably larger down-type quark and lepton masses. 

The distributions for the various masses are shown in figure~\ref{fig:scan_masses_MSSM} compared to their observed values. As anticipated, the model, qualitatively, is remarkably successful. A fully hierarchical spectrum, with significant 1-2 and 2-3 splittings, is a typical outcome of this framework.
As expected, the muon mass prediction can be satisfied only at the extreme tail of our distribution, a problem we will address later.

It is also enlightening to inspect the two-dimensional distribution of first and second generation masses. 
This unveils any non-trivial correlations among these parameters. A scatter of scan points is shown in figure~\ref{fig:1st2nd_MSSM} for the up-type and down-type quarks as well as for leptons. We see that the up-type sector is in good agreement with observation. For the down-type sector, observation is at the edge of the model prediction and we would benefit from increasing the down type mass. 
In the lepton sector, the situation can improve by increasing both muon and electron masses, while keeping their ratio fixed.

\section{Quark Mixing}\label{sec:mixing}

We now turn to examining quark mixing, i.e. the misalignment of the left-handed up and down quarks in flavor space. We will find that in the minimal model quark mixing is generated radiatively, but that the Cabibbo angle is typically too small. We will then show that this can be addressed simply by allowing for flavor violating trilinear terms.

\subsection{Quark Mixing with Flavor Aligned Trilinears} \label{sec:CKM}

Just like fermion masses, the CKM matrix is generated radiatively in our framework. We will first analyze quark mixing in the minimal model as it has been described so far. In line with the previous subsections, we will 
first get a qualitative understanding of the result, and then proceed to consider the results analytically (in the mass insertion approximation) and numerically (without approximation).

As was the case for the fermion masses, it is enlightening to inspect the loop induced contribution to the CKM matrix at the \emph{integrand} level, because symmetries that are exact for the integrand are manifested approximately at the integral level due to the dominance of a single scale. Purposely dropping the loop integral, the up-type and down-type mass matrices have the following structure
\begin{equation}
\label{eq:up-and-down}
m^{ij}_u \sim m^{ij}_{\mathrm{tree}} + f^i_{q_L} f^j_{u_R}  ~, \qquad\qquad
m^{ij}_d \sim m^{ij}_{\mathrm{tree}} + f^i_{q_L} f^j_{d_R} ~.   
\end{equation}
To see the parametric structure of the CKM matrix we can pick a convenient basis. Obviously, we will want to work in a basis where only the 33 entry of $m_\mathrm{tree}$ is non-zero. We can further do $U(2)$ rotations between the lighter generations such that $f^1_{q_L}=f^1_{d_R}=f^1_{u_R}=0$. In this basis both the up and the down matrices in equation~(\ref{eq:up-and-down}) are non-vanishing only in the lower two-by-two diagonal block. These matrices can thus be diagonalized with a rotation between the second and third generation. The relative left handed rotation between them, which is the CKM matrix, will also be purely a 2-3 rotation. Recalling that the structure of equation~(\ref{eq:up-and-down}) is merely approximate, we expect a suppression of the 1-2 and 1-3 CKM entries.

The CKM matrix is fully determined by four parameters. We chose the Cabibbo angle $\lambda_C$ and the absolute values of the CKM elements $V_{cb}$ and $V_{ub}$ as well as the CKM phase. Our expectation from above is that $V_{ub}$ and $\lambda_C\equiv V_{us}$ will be suppressed. Using the same approximations as in section~\ref{sec:minimal_analytic}, we find the following analytic expressions
\begin{subequations}
\begin{eqnarray} \label{eq:lambda}
 \lambda_C &\simeq& \frac{3}{125} ~|\delta^L_{12}| \left( |\delta^R_{cu}|^2 - |\delta^R_{sd}|^2 \right)  ~,\\
 |V_{cb}| &\simeq& \frac{\alpha_s}{4\pi} \frac{4}{9} ~|\delta^L_{23}|~ \frac{|m_{\tilde g} \mu|}{m_{\tilde q}^2} \frac{t_\beta}{|1 + \epsilon_b t_\beta|} ~, \label{eq:Vcb} \\
 |V_{ub}| &\simeq& \frac{\alpha_s}{4\pi} \frac{2}{45} ~|\delta^L_{12}\delta^L_{23}|~ \frac{|m_{\tilde g} \mu|}{m_{\tilde q}^2} \frac{t_\beta}{|1 + \epsilon_b t_\beta|} ~. \label{eq:Vub}
\end{eqnarray}
\end{subequations}
These expressions have to be compared to the measured values that are collected in the appendix~\ref{app:masses}.
The physical phase of the CKM matrix is generically of $O(1)$ in our model, in agreement with observation. The smallness of the CKM element $V_{cb}$ is explained by a loop factor. Its measured value can be easily accommodated in the large $\tan\beta$ regime. As explained above, the loop induced quark mass matrices are approximately rank 1, and mixing between the first and third generation is therefore suppressed by a small numerical factor. The resulting expression for the CKM element $V_{ub}$ is smaller than $V_{cb}$ by one order of magnitude, exactly as observed.
Finally, the Cabibbo angle $\lambda_C$ is not loop suppressed because it parametrizes the mixing between the first and second generation of quarks, which both acquire mass at the same loop level. However, again due to the fact that the loop induced quark mass matrices are approximately rank 1, the Cabibbo angle is suppressed by a small numerical factor. Moreover, the mixing between the first and second generation of left handed quarks is proportional to $\delta_{12}^L$ and aligned in the up and down sectors at leading order in mass insertions. The Cabibbo angle is therefore only induced at higher order in the mass insertion approximation, involving also the mixing of right handed quarks. This double suppression leads to a very small value of $\lambda_C$. Even setting the mass insertions in~(\ref{eq:lambda}) to 1, $\lambda_C$ is almost one order of magnitude below the observed value.

\begin{figure}[t]
\centering
\includegraphics[width=0.43\textwidth]{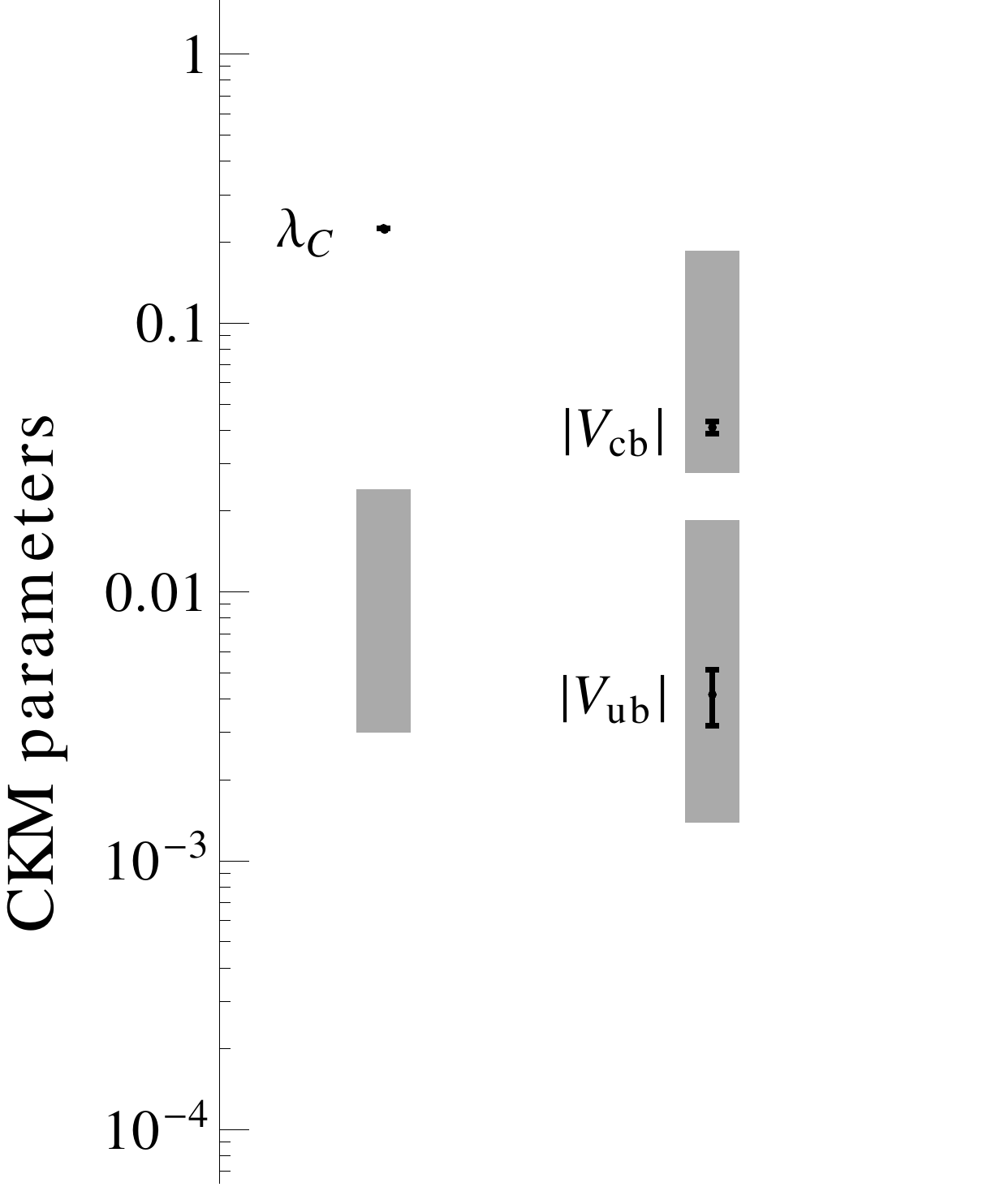} \qquad
\includegraphics[width=0.43\textwidth]{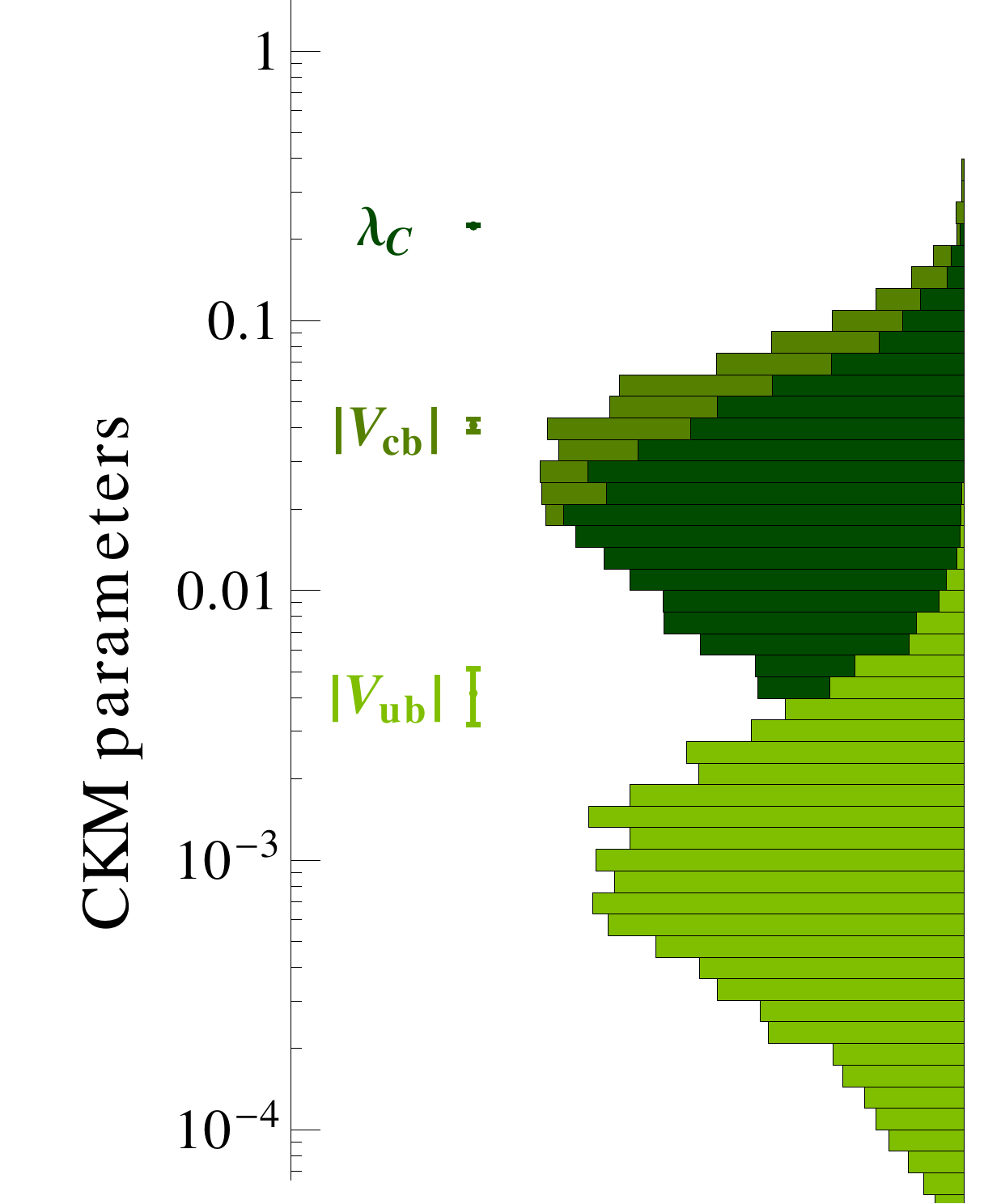} 
\caption{
(Left): Generic predictions for the structure of the CKM matrix elements, analogous to the mass spectrum shown in figure~\ref{fig:chart_MSSM}.
(Right): Distributions of the CKM parameters as obtained by a parameter scan of the minimal model. Shown are also the observed values of the CKM parameters with $2\sigma$ uncertainties.}
\label{fig:scan_CKM_MSSM}
\end{figure}

The generic predictions for the CKM parameters~(\ref{eq:lambda}), ~(\ref{eq:Vcb}), and~(\ref{eq:Vub}) are also summarized in the left plot of figure~\ref{fig:scan_CKM_MSSM} where we set all mass scales equal and vary $0.5 < \delta^A_f < 1$ and $30 < \tan\beta < 100$.
In the right plot of figure~\ref{fig:scan_CKM_MSSM}, we show the distributions of the CKM parameters that result from the parameter scan discussed in section~\ref{sec:minimal_numeric}. The observed values for $|V_{cb}|$ and $|V_{ub}|$ can be easily reproduced by the model. However, as expected from the analytical results, the Cabibbo angle is generically one order of magnitude below the measured value and tends to be even smaller than $|V_{cb}|$. The next section will describe ways to increase the Cabibbo angle.

\subsection{Flavor Violating $A$-terms} \label{sec:Aterms}

We now consider the possibility that all soft terms, both masses and trilinears, violate the flavor symmetry. Traditionally, in supersymmetric theories the $A$-terms are taken to be proportional to the Yukawa matrices for several reasons - constraints on flavor changing neutral currents, vacuum stability, as well as model building considerations. Flavor alignment is certainly motivated from a model building point of view but is not necessary - one could imagine writing an operator of the form 
\begin{equation}
\label{eq:FVaterm}
\mathcal{L}_\text{soft} \supset \int d^2\theta \frac{X}{M}\lambda_{ij}H_uQ_iU_j ~,
\end{equation}
where $X$ is a hidden sector field that gets an $F$-term vev and $M$ is a messenger scale. The matrix $\lambda$ need not be proportional to the Yukawa matrix. Indeed, in gravity mediation, where $M$ is the Planck scale, the lore is that the theory of quantum gravity does not respect any flavor symmetries and one could naively expect $\lambda_{ij}$ in equation~(\ref{eq:FVaterm}) to be not aligned with the Yukawa matrix.

The constraints on flavor violating $A$-terms from flavor changing neutral currents are avoided in our framework due to the large supersymmetry breaking scale, but constraints from vacuum stability certainly do need to be considered. Trilinear scalar terms can always drive an instability at large field values unless they are countered by a sufficiently large quadratic term. In the MSSM the supersymmetric Yukawa terms serve as stabilizing quartic couplings. If the $A$-terms are significantly not aligned with these Yukawas, a runaway direction may arise. Completely anarchic trilinear couplings are therefore subject to strong constraints from vacuum stability~\cite{Casas:1996de,ArkaniHamed:1996zw,Borzumati:1999sp}. In particular, soft trilinear couplings of the first and second generation as well as trilinears that mix the first with the second generation are strongly constrained, because F-term quartics for the first and second generation sfermions are absent at tree level in the considered framework. Flavor changing trilinear couplings involving the third generation are less constrained because they are at least partially aligned with the Yukawa couplings. 

Focusing on flavor changing trilinears in the up sector, we write
\begin{equation}
 \mathcal{L}_\text{soft} \supset Y_t A_{ut} H_u \tilde q_1 \tilde u_3 + Y_t A_{ct} H_u \tilde q_2 \tilde u_3 + Y_t A_{tu} H_u \tilde q_3 \tilde u_1 + Y_t A_{tc} H_u \tilde q_3 \tilde u_2 + \mathrm{h.c.} ~.
\end{equation}
Necessary bounds on the flavor changing trilinears in the up sector read~\cite{Casas:1996de}
\begin{subequations}
\begin{eqnarray} \label{eq:FCAbound}
 |A_{ut}|^2 \lesssim m_{\tilde u_L}^2 + m_{\tilde t_R}^2 ~&,&~~ |A_{ct}|^2 \lesssim m_{\tilde c_L}^2 + m_{\tilde t_R}^2 ~, \\
 |A_{tu}|^2 \lesssim m_{\tilde t_L}^2 + m_{\tilde u_R}^2 ~&,&~~ |A_{tc}|^2 \lesssim m_{\tilde t_L}^2 + m_{\tilde u_R}^2 ~.
\end{eqnarray}
\end{subequations}
In general it is possible to derive stronger bounds by optimizing the direction in field space.
Limiting ourselves to the case discussed below, namely the presence of $A_t$, $A_{ut}$ and $A_{tu}$ trilinear couplings, and assuming for simplicity flavor diagonal soft masses, we find the following bound
\begin{equation} \label{eq:Atubound}
 \left( |A_t| + t_L |A_{ut}| + t_R |A_{tu}|\right)^2 \lesssim \left(3+ t_L^2 + t_R^2 \right) \left( c_L^2 m_{\tilde t_L}^2 + s_L^2 m_{\tilde u_L}^2 + c_R^2 m_{\tilde t_R}^2 + s_R^2 m_{\tilde u_R}^2 \right) ~,
\end{equation}
with $t_{L/R} = \tan\theta_{L/R}$, $s_{L/R} = \sin\theta_{L/R}$, and $c_{L/R} = \cos\theta_{L/R}$. The inequality in~(\ref{eq:Atubound}) has to hold for all angles $\theta_{L/R}$. In the limit $\theta_L = \theta_R = 0$ we recover the bound on $A_t$ given in~(\ref{eq:Atbound}); for $\theta_L = \pi/2$ and $\theta_R = 0$ we obtain the first bound in equation~(\ref{eq:FCAbound}), etc.
Similar to the case of the flavor conserving trilinears discussed in section~\ref{sec:stability}, the vacuum stability bounds can become tighter, if one takes into account also flavor violation in the soft masses. The bounds will be slightly relaxed if one demands only meta-stability~\cite{Park:2010wf}.

Our philosophy for flavor violating $A$-terms will be similar to flavor violating soft masses: we will allow anarchic flavor violation so long as it does not cause a vacuum instability. For the soft masses this amounted to checking that there are no tachyons in the spectrum, but for trilinears the requirement is that no runaway or instability is caused at large field values. In practice this implies that only trilinears that involve the third family can be turned on.

In the following, we consider the effect of switching on a small flavor changing $A_{ut}$ and~$A_{tu}$ trilinear coupling. It is well known that flavor changing trilinear couplings can lead to large corrections to the CKM matrix elements~\cite{Hamzaoui:1998yy,Crivellin:2008mq}. 
In general we expect this new source of flavor violation to mildly spoil the approximate rank 1 structure described in section~\ref{sec:mechanism}. We thus expect the up quark mass, as well as $\lambda_C$ and $V_{ub}$ to be particularly sensitive to the flavor changing trilinears.
To understand the dependence of these parameters on the flavor violating $A$-terms we employ the mass insertion approximation defined in section~\ref{sec:minimal_analytic}, working to leading order in $A_{ut}/A_t$ and $A_{tu}/A_t$. 
We further assume that $A_{ut}/A_t$ and $A_{tu}/A_t$ are of order $\delta^2$ (smaller than $\delta$). We find the following CKM matrix elements 
\begin{subequations}
\begin{eqnarray} \label{eq:lambda_2}
 \lambda_C &\simeq& \frac{2 |A_{ut}|}{|\delta^L_{23} A_t|} ~,\\
 |V_{cb}| &\simeq& \frac{\alpha_s}{4\pi} \frac{4}{9} ~|\delta^L_{23}|~ \frac{|m_{\tilde g} \mu|}{m_{\tilde q}^2} \frac{t_\beta}{|1 + \epsilon_b t_\beta|} ~,\\
 |V_{ub}| &\simeq& \frac{\alpha_s}{4\pi} \frac{2}{45} ~\left|\delta^L_{12}\delta^L_{23}+ 20 \frac{A_{ut}}{A_t}\right|~ \frac{|m_{\tilde g} \mu|}{m_{\tilde q}^2} \frac{t_\beta}{|1 + \epsilon_b t_\beta|} ~.
\end{eqnarray}
\end{subequations}
At the considered order, only the flavor changing trilinear $A_{ut}$ affects the CKM elements. Values for $A_{ut}$ approximately one order of magnitude smaller than $A_t$, $|A_{ut}| \sim 0.1 \times |A_t|$, are sufficient to generate the observed Cabibbo angle. An $A_{ut}$ of that size also leads to an $O(1)$ shift in the CKM element $V_{ub}$.
Note that the size of $A_{ut}$, required to generate the observed Cabibbo angle is well below the vacuum stability bound in equation~(\ref{eq:FCAbound}).
Both flavor changing trilinears, $A_{ut}$ and $A_{tu}$ affect the up quark mass
\begin{eqnarray} \label{eq:up_mass}
\frac{m_u}{m_t} &\simeq& \frac{\alpha_s}{4\pi} \frac{2}{225} \frac{|m_{\tilde g} A_t|}{m_{\tilde q}^2} \left|\delta_{12}^L\delta_{23}^L\delta_{tc}^R\delta_{cu}^R -5 \left(\delta_{12}^L\delta_{23}^L \frac{A_{tu}}{A_t} + \frac{A_{ut}}{A_t} \delta_{tc}^R\delta_{cu}^R \right) \right|~.
\end{eqnarray}
For $|A_{ut}|,|A_{tu}| \sim 0.1 \times |A_t| $, the shift in the up quark mass is of $O(1)$ and therefore does not change the predicted hierarchy between up and charm mass significantly.

\begin{figure}[t]
\centering
\includegraphics[width=0.43\textwidth]{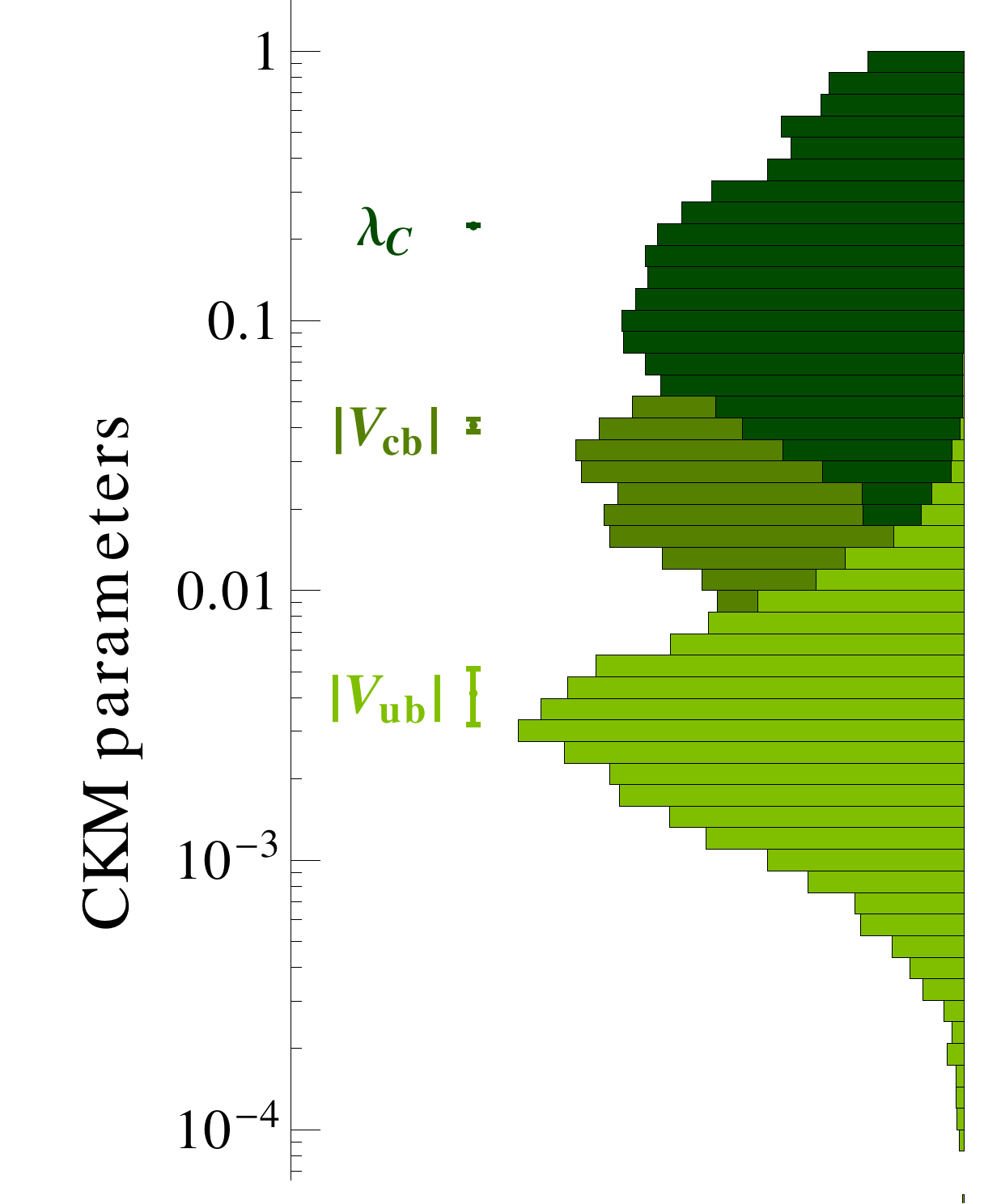}
\caption{
Distributions of the CKM parameters as obtained by a parameter scan of the minimal model, including a holomorphic trilinear coupling $A_{ut}$. Shown are also the observed values with $2\sigma$ uncertainties.
}
\label{fig:scan_FVaterms}
\end{figure}

Results for the CKM parameters coming from a numerical scan that includes a non-zero $A_{ut}$ are shown in figure~\ref{fig:scan_FVaterms}. We randomly scan $0.01 \times \tilde m < |A_{ut}| < 0.1 \times \tilde m$ (with a flat distribution) and allow for an arbitrary complex phase of $A_{ut}$. The agreement between the predicted values for the CKM elements and the observed values is greatly improved compared to the scan without flavor changing trilinears.

Flavor violating trilinear couplings in the down sector can lead to additional contributions to the down mass matrix and in turn also affect the CKM matrix. However, these contributions are suppressed by $1/\tan\beta$ with respect to the leading contributions that are induced by the flavor conserving supersymmetric trilinear couplings proportional to the $\mu$ term~(\ref{eq:mutrilinear}).
As a consequence, effects from flavor violating trilinears in the down sector are insignificant.

Another possibility would be to introduce flavor violating trilinear couplings in the down sector that are non-holomorphic and therefore not suppressed by $1/\tan\beta$. A non-holomorphic $A^\prime_{db}$ coupling with absolute values approximately 10\% of the $\mu$ term would have similar effects on the CKM matrix as $A_{ut}$ discussed above.
A non-holomorphic $A^\prime_{bd}$ coupling of the same order as the $\mu$ term could be used to enhance the down quark mass by a factor of few without affecting the CKM matrix significantly. 
However, flavor violating non-holomorphic A-terms in the down sector are strongly constrained by vacuum stability considerations. We find that with trilinear couplings of the type $A^\prime_{db} \tilde q_1 \tilde d_3 H_u^*$ and $A^\prime_{bd} \tilde q_3 \tilde d_1 H_u^*$, the MSSM tree level potential is unbounded from below. The electro-weak minimum can still be metastable in such a situation~\cite{Borzumati:1999sp}, but a dedicated analysis to establish the maximal values for $A^\prime_{bd}$ and $A^\prime_{db}$ compatible with meta-stability bounds is beyond the scope of this work.

\section{Modifications for a Milder Mass Hierarchy}\label{sec:mod}

In section~\ref{sec:spectrum} we saw that our minimal setup, the MSSM with third generation Yukawas and anarchic soft masses, successfully generates a hierarchical fermion spectrum. When compared to the observed hierarchy we find that, if anything, the hierarchy is too large. In particular, the ratio of the muon to tau mass is predicted to be too small. In this section we will consider variations to the minimal framework which yield a milder hierarchy. The first possibility we consider, still within the minimal model, is to allow for hard supersymmetry breaking. The second alternative we will discuss, is to extend the gauge structure beyond that of the MSSM.
 
\subsection{Hard Supersymmetry Breaking}\label{sec:hard}

So far we have considered a scaled up version of the traditional softly broken supersymmetric standard model. Introducing only soft SUSY breaking is certainly well motivated by naturalness. But given that we have given up on electro-weak naturalness in the first paragraph of this paper, it is reasonable to consider also hard breaking of supersymmetry. Here we define hard breaking as any symmetry breaking by a dimensionless coupling. For practical purposes we will consider such breaking of supersymmetry as a limit of soft supersymmetry breaking models by taking the supersymmetry breaking vev $F$ to be of the same order as the messenger scale $M^2$. Then, dimensionless SUSY breaking operators can be categorized in powers of $F/M^2$~\cite{Martin:1999hc}.

For our purposes, we will be interested in the possibility that the bino-fermion-sfermion vertex is larger than the hypercharge gauge coupling $g_1$. Within our framework this can come from the operator
\begin{equation}
\label{eq:hard}
\mathcal{L}\supset \int d^4\theta \frac{X^\dagger X}{M^4} \mathcal{W}^\alpha\, \Phi^\dagger D_\alpha \Phi + \mathrm{h.c.}= 
 \frac{F^2}{M^4} \lambda^\alpha \phi^* \psi_\alpha + \mathrm{h.c.} ~.
\end{equation}
We will assume that this new source of supersymmetry breaking is flavor blind, allowing the main mechanisms we have discussed to remain intact. We note that relaxing this assumption somewhat may be interesting as an alternative to the flavor violating trilinears discussed in subsection~\ref{sec:Aterms}.
The scale $F$ may be the only source of supersymmetry breaking, in which case \emph{the} mediation mechanism may be low scale. Alternatively, there can be two sources of SUSY breaking giving two different scales $F_\mathrm{high}$ and $F_\mathrm{low}$. The scale $F_\mathrm{high}$ can be coupled to the MSSM via gravity mediation giving rise to the lions share of soft masses. The scale $F_\mathrm{low}$ can be coupled via a low scale mediation, $F\sim M^2$, which contributes up to an order one of the soft masses as well as some hard breaking of the form of equation~(\ref{eq:hard}). A setup like this fits our framework well because the high-scale gravity mediation is expected to be flavor anarchic while the low scale mediation could be flavor blind. 

\begin{figure}[t]
\centering
\includegraphics[width=0.45\textwidth]{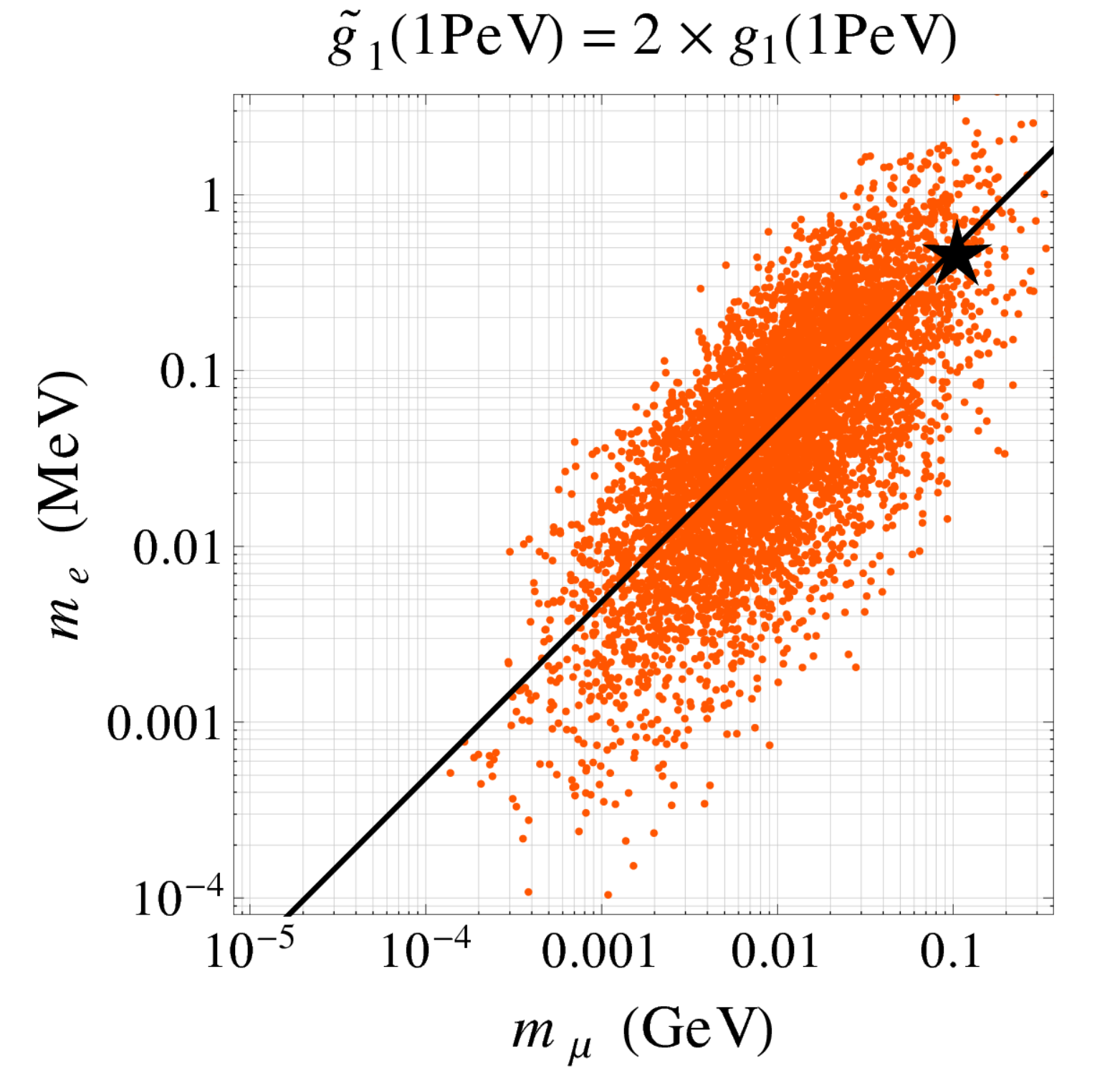} ~~~~~
\includegraphics[width=0.45\textwidth]{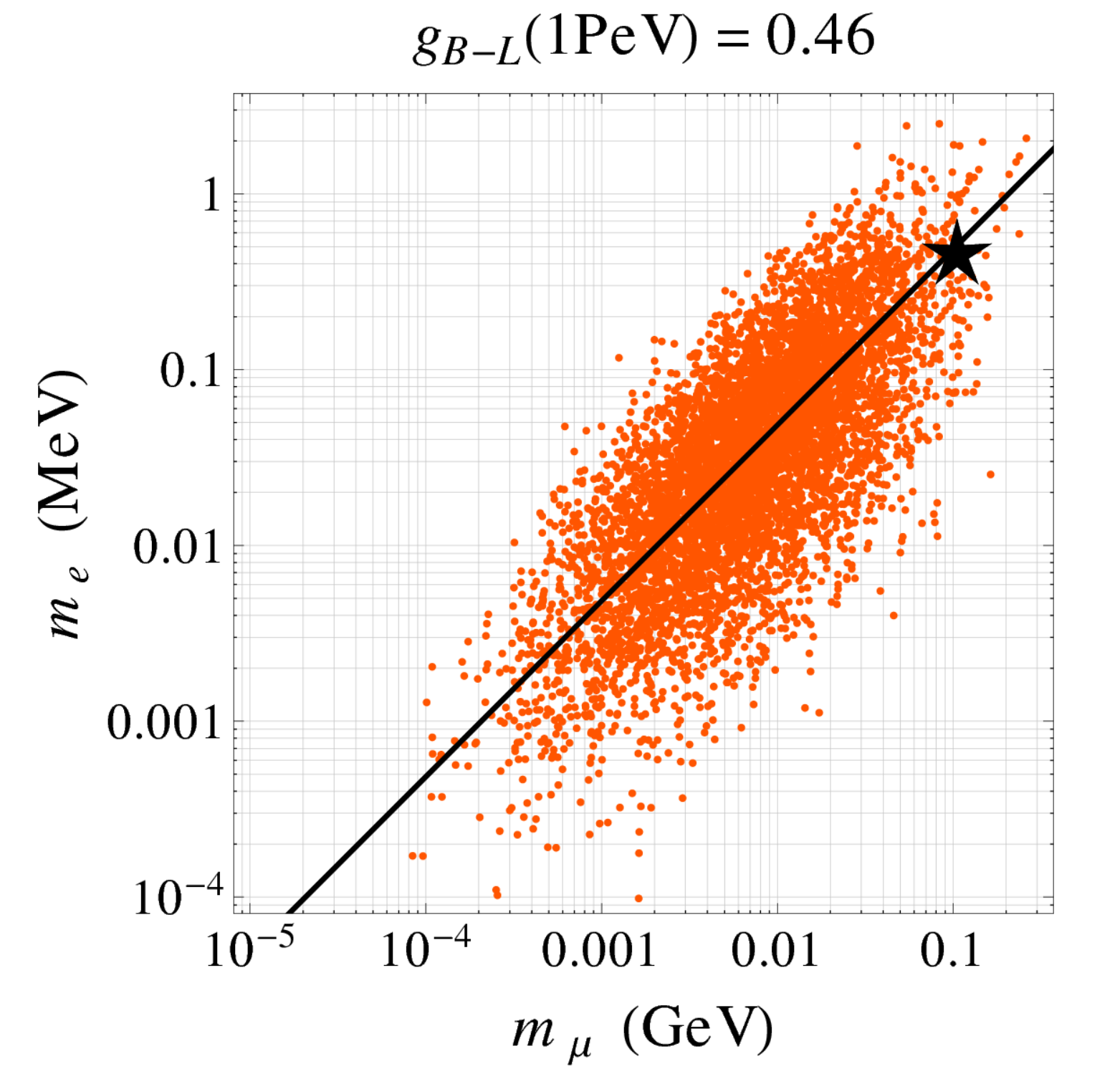}
\caption{The lepton mass spectrum may be improved in variations of the minimal model. We show parameter scans for the electron and muon masses, \emph{(left)} in models where hard supersymmetry breaking is introduced, and \emph{(right)} in models with a gauged $B-L$ number. In both cases the agreement may be improved by increasing the relevant couplings, see text for details.}
\label{fig:hard_BL}
\end{figure}

Increasing the bino-fermion-sfermion coupling $\tilde g_1$ from its supersymmetric value $g_1$, will modify the fermion spectrum in our model. In particular, both the electron and muon masses will be shifted by a factor of $(\tilde g_1/g_1)^2$. This will improve the agreement of these masses with observation significantly as can be seen in the left panel of figure~\ref{fig:hard_BL} where we have taken $\tilde g_1$ to be twice its supersymmetric value. This distribution is similar to that in figure~\ref{fig:1st2nd_MSSM} with all points shifted by a factor of four along the diagonal line. We note that introducing significant hard SUSY breaking generically also leads to larger radiative corrections to third generation masses (the parameters we called $\epsilon_\tau$).

Increasing $\tilde g_1$ does not affect the quark masses significantly as these were dominated by squark-gluino loops. Increasing the quark masses with hard SUSY breaking is possible for example by increasing the quark-squark-gluino vertex $\tilde g_3$ in a similar way. Increasing the first and second generation masses by a common factor does indeed bring the bulk of our scan models closer to observation. However we note that the observed quark masses are certainly achievable within our minimal framework.

Finally, we note that hard supersymmetry breaking may affect the Higgs mass. As discussed in Appendix~\ref{app:higgs}, our frameworks requires lowering the Higgs mass compared to the standard supersymmetric prediction. We discuss this further in that Appendix.

\subsection{Beyond The MSSM: Extended Gauge Structures}\label{sec:B-L}

We now return to the framework of traditional softly broken supersymmetry and consider and alternative solution to raising the muon mass. In our minimal framework the mass ratio of second to third generation fermions is a function of the gauge representations and the MSSM gauge couplings. A simple possibility to generating a milder hierarchy in the lepton sector is to assume that additional gauge symmetries are present at the supersymmetry breaking scale. Indeed, decades of studying TeV scale supersymmetry has taught us that $\tilde m_\mathrm{SUSY}$ is a natural scale for spontaneously broken gauge symmetries. Any additional gauge symmetry under which both left and right handed matter transforms will include new diagrams similar to those in figure~\ref{fig:diagrams1} but with a new gaugino. 

There are several ways one could envision enlarging the gauge structure of the MSSM and we will leave a more complete analysis of the most motivated possibilities for future study. Here, as a simple example we consider gauging $B-L$ in addition to the SM gauge group. The $B-L$ gauge symmetry can naturally be spontaneously broken at the scale of supersymmetry breaking $\tilde m$. Due to their larger charge, lepton masses would be affected by the new $B-L$ gaugino diagrams more than quarks. As opposed to the case of hard SUSY breaking, where we needed to assume the new contribution is flavor universal, in this case flavor universality is ensured by $B-L$ gauge invariance and the hierarchical structure of section~\ref{sec:mechanism} is automatically maintained.    
In the right panel of figure~\ref{fig:hard_BL} we show the effects of including a $B-L$ gauge interaction whose coupling strength is $g_{B-L} = 0.46$ at the PeV scale. This value is chosen by requiring that the Landau pole in the $B-L$ coupling is postponed until the GUT scale, thus maintaining the success of gauge coupling unification\footnote{A full analysis of unification, which we do not perform here, would need to consider the kinetic mixing between $B-L$ and hypercharge that is induced by RGE's~\cite{Babu:1996vt} or add additional matter that renders the two sets of charges orthogonal.}. Comparing this to figure~\ref{fig:1st2nd_MSSM}, we conclude that gauging $B-L$ can bring the value of lepton masses within the range predicted by our framework. 

Of course, one can bring the most ``likely'' values of the muon mass closer to the observed value by increasing the $B-L$ coupling. This may be done at the cost of introducing a Landau pole below the the canonical GUT scale. Unification of gauge couplings would still be possible in this case if it is accelerated~\cite{ArkaniHamed:2001vr}. Alternatively, one could embed the $B-L$ into a larger non-abelian group which is asymptotically free before the Landau pole. 

We reiterate that $B-L$ is brought here as an example of how adding gauge structure around the supersymmetry breaking scale can improve our predicted lepton spectrum. Turning this statement around we can consider the degree of hierarchy in the various SM representations as an indication of their gauge charges at~$\tilde m$. Taking this seriously may suggest a gauge symmetry under which leptons are charged fairly strongly, such as Pati-Salam 
models~\cite{Pati:1974yy}. We leave this line of inquiry for an upcoming publication.

\section{Unification} \label{sec:unification}

\begin{figure}[t]
\centering
\includegraphics[width=0.45\textwidth]{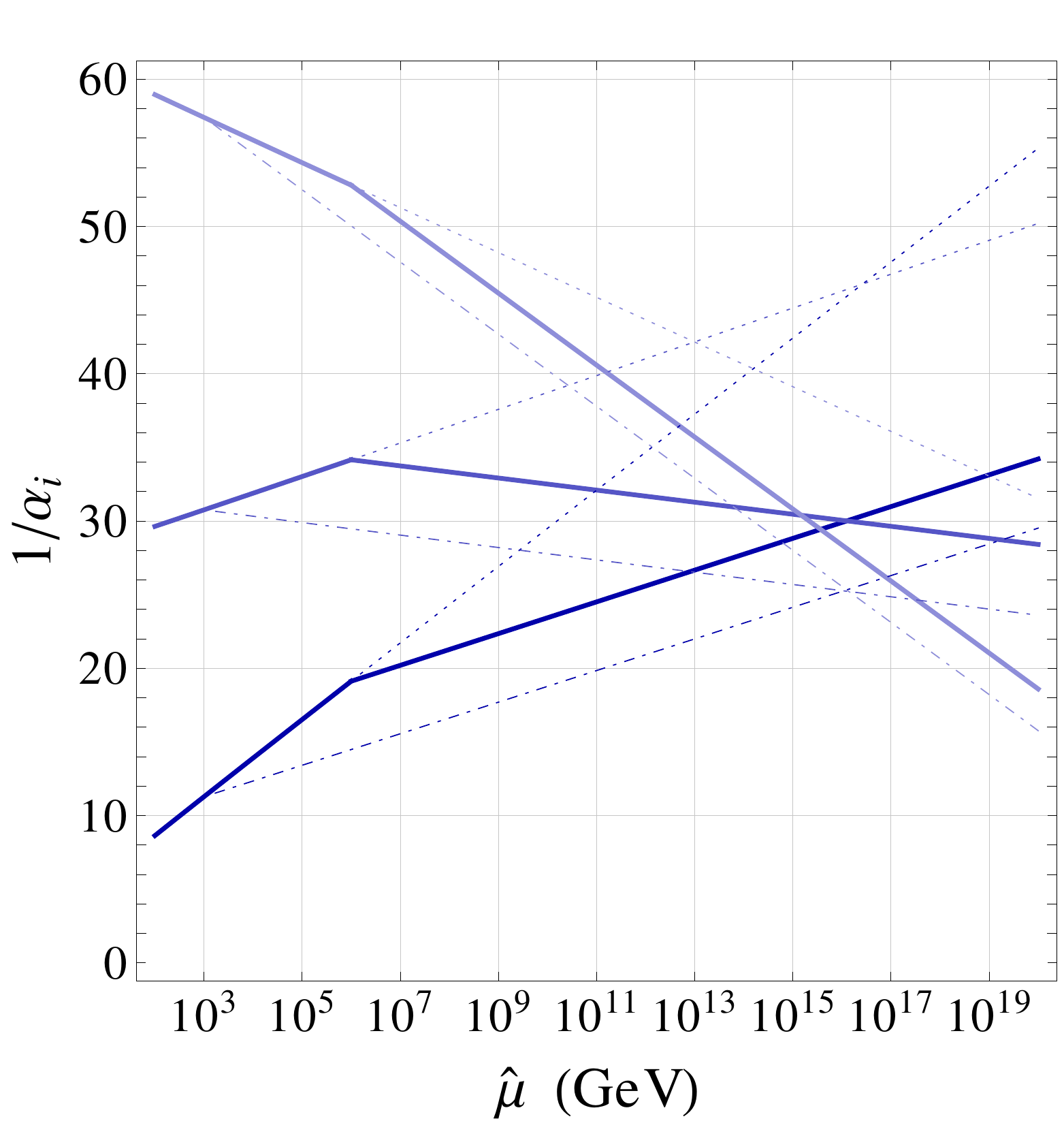}
\caption{Running of gauge couplings with the MSSM matter content with a SUSY breaking scale of~1~PeV (solid). For comparison the SM running (dotted) and the running in the MSSM with TeV SUSY breaking (dash-dot) are also shown.}
\label{fig:gaugeunification}
\end{figure}

We now discuss the compatibility of our framework with unification of gauge and Yukawa couplings. The discussion here applies to the minimal model and its extension in section~\ref{sec:hard}. As stated above, the the compatibility of the $B-L$ extension of section~\ref{sec:B-L} warrants further study.   
One of the attractive features of TeV scale SUSY is that it is compatible with gauge coupling unification at a GUT scale around $2\times 10^{16}$ GeV~\cite{Dimopoulos:1981zb, Dimopoulos:1981yj, Amaldi:1991cn,Langacker:1991an}. Given the existing constraints on flavor changing neutral current processes, the anarchic scalar masses in our framework must be of order 1~PeV or above. As is shown in figure~\ref{fig:gaugeunification}, gauge coupling unification is still compatible with a SUSY breaking scale of a PeV. For the running of the gauge couplings we use 2-loop RGEs. Up to the PeV scale we use the SM RGEs~\cite{Machacek:1983fi,Machacek:1983tz,Machacek:1984zw,Luo:2002ey} and above the PeV scale we use those of the MSSM~\cite{Martin:1993zk}. We neglect threshold corrections at the SUSY scale. In that case, threshold corrections at the GUT scale at the level of 1\% are required for successful gauge coupling unification (see also~\cite{Bagnaschi:2014rsa}). 

Besides gauge coupling unification, SUSY GUTs can also be compatible with third generation Yukawa unification~\cite{Hall:1993gn,Carena:1994bv,Blazek:2002ta,Altmannshofer:2008vr}, e.g. bottom-tau Yukawa unification in GUTs based on the $SU(5)$ gauge group, or top-bottom-tau Yukawa unification in $SO(10)$ GUTs. We find that our framework can accommodate Yukawa unification easily. To show this we run the third generation Yukawa couplings to the GUT scale using 2-loop RGEs. 
At the SUSY scale we incorporate threshold corrections to the bottom and tau Yukawas, $\delta_b$ and $\delta_\tau$,
\begin{equation}
 y_t^\text{SUSY} = y_t^\text{SM} / s_\beta ~,~~ y_b^\text{SUSY} = y_b^\text{SM} (1 + \delta_b)  / c_\beta ~,~~ y_\tau^\text{SUSY} = y_\tau^\text{SM} (1 + \delta_\tau)  / c_\beta ~.
\end{equation}
Threshold corrections to the top Yukawa coupling are at most at the percent level and therefore negligible.

%
\begin{figure}[t]
\centering
\includegraphics[width=0.45\textwidth]{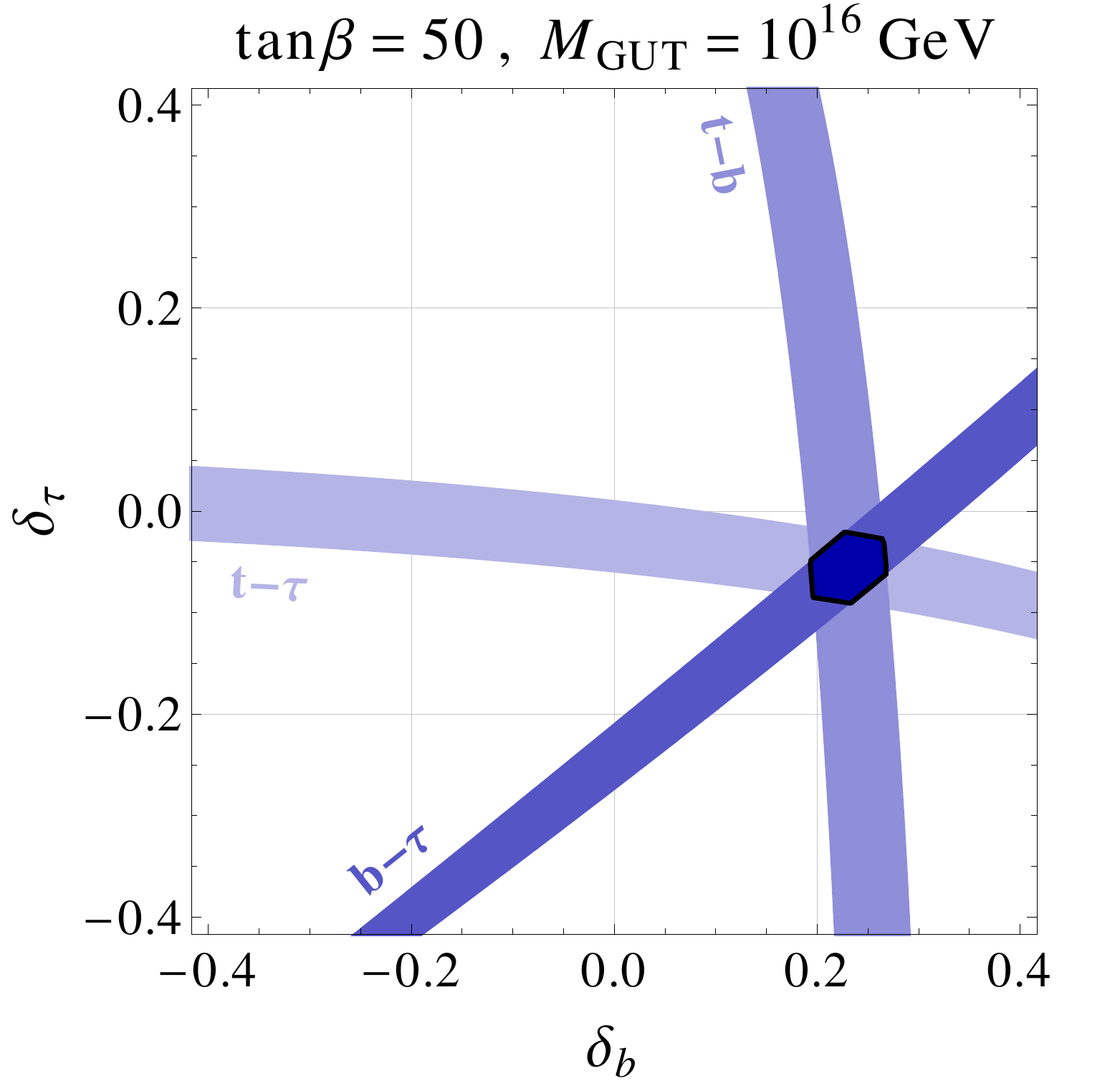} ~~~~~
\includegraphics[width=0.45\textwidth]{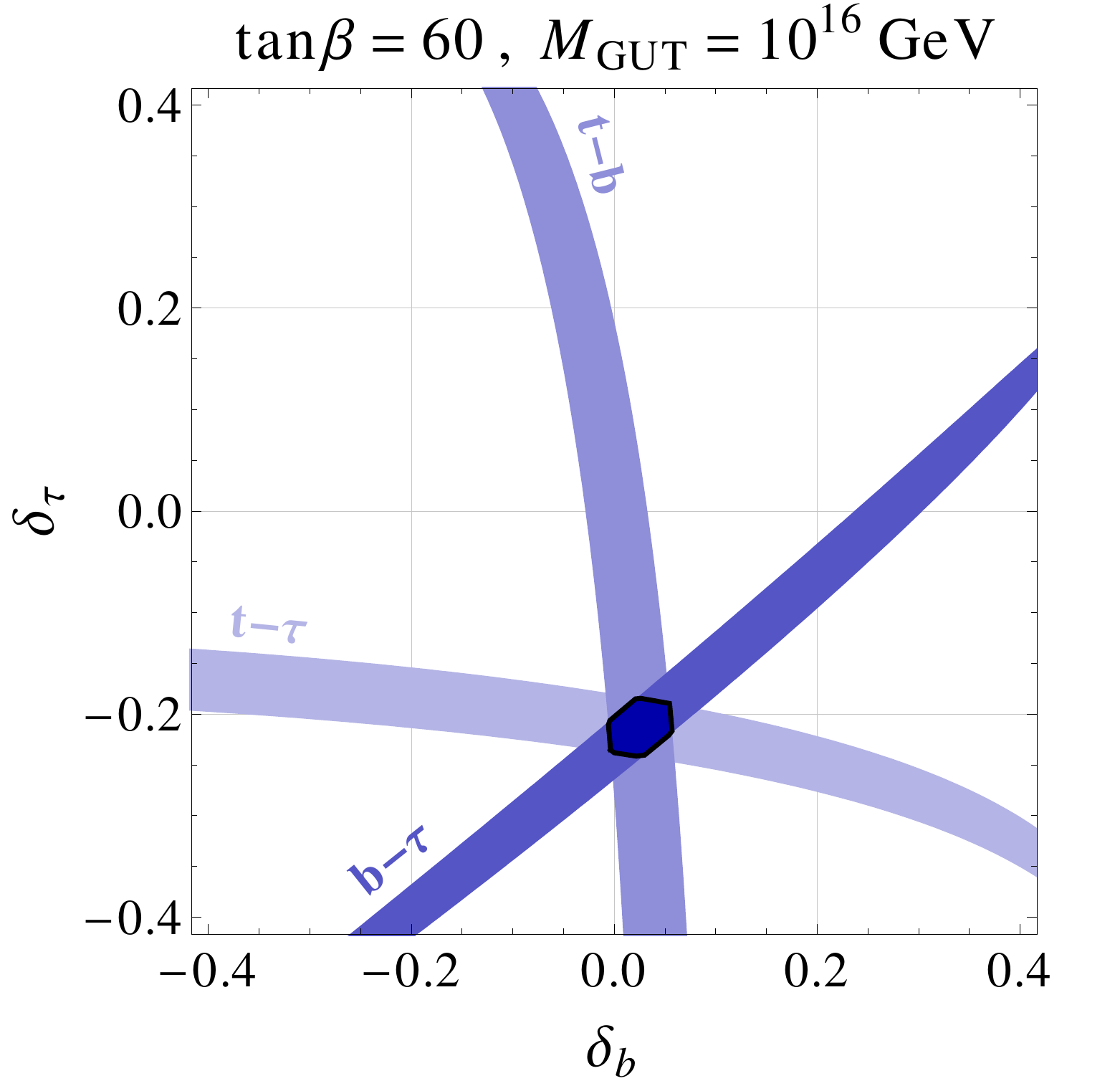}
\caption{Threshold corrections to the bottom and tau Yukawa couplings at a SUSY scale of 1~PeV that are required for third generation Yukawa unification at a GUT scale of $10^{16}$~GeV. In the left (right) plot we set $\tan\beta = 50$ ($\tan\beta = 60$).}
\label{fig:yukawaunification}
\end{figure}
%
In figure~\ref{fig:yukawaunification} we show the size of the threshold corrections to the bottom and tau Yukawa couplings required to achieve third generation Yukawa unification at a scale of $10^{16}$~GeV. 
These are shown for $\tan\beta = 50$ and $\tan\beta = 60$. The shaded bands correspond to the regions that are compatible with top-bottom, top-tau, and bottom-tau unification with an accuracy of better than 5\%.
The threshold corrections required for Yukawa unification can be easily accommodated in our framework.
They are approximately given by
\begin{equation}
 \delta_b \simeq \frac{1}{|1 + \epsilon_b t_\beta|} - 1 ~,~~ \delta_\tau \simeq \frac{1}{|1 + \epsilon_\tau t_\beta|} - 1 ~,
\end{equation}
with $\epsilon_b$ and $\epsilon_\tau$ given in~(\ref{eq:eps}).
Note that requiring third generation Yukawa unification does not spoil the successful generation of the first and second generation quark and lepton masses. We checked explicitly that enforcing bottom-tau unification by requiring $\delta_b$ and $\delta_\tau$ to lie within the corresponding band shown in figure~\ref{fig:yukawaunification}, does not change the distribution of the radiatively induced fermion masses and quark mixing angles appreciably.

\section{Summary and Outlook} \label{sec:conclusion}

Models of radiative fermion masses are an elegant and attractive possibility to explain the hierarchical flavor pattern in the mass spectrum and mixing angles of the quarks and leptons in terms of loop factors.
In this work, we discussed setups that allow to generate the hierarchical flavor structure of the SM fermions from loops of superpartners with anarchical soft masses. 
Given the anarchic squark and slepton masses, the discussed setups are subject to strong constraints from low energy flavor observables. In particular, CP violation in Kaon mixing generically sets the strongest lower bound on the SUSY scale of the order of $m_\text{SUSY} \gtrsim 1$~PeV. 
A spectrum with all SUSY particles at the PeV scale is still compatible with gauge coupling unification and also with third generation Yukawa unification at a GUT scale of around $m_\text{GUT} \sim 10^{16}$~GeV. 

We found that the Minimal Supersymmetric Standard Model with flavor anarchic soft masses for the squarks and sleptons contains all necessary ingredients to generate a fully hierarchical spectrum of quarks and leptons. At tree level, rank 1 Yukawas couple the Higgs only to the third generation fermions and leave a $U(2)^5$ symmetry acting on the first and second generation unbroken. The anarchic sfermion soft masses, on the other hand, break the SM flavor symmetry completely. Loops containing gauginos and sfermions communicate the flavor symmetry breaking from the sfermion sector in the fermion sector and
Yukawa couplings for the first and second generation fermions are generated at the 1-loop level.
While both first and second generation Yukawas are generated at the 1-loop level, we pointed out a novel mechanism that ensures a large splitting, of generically two orders of magnitude, without introducing small parameters or symmetries. We showed that the radiatively induced Yukawa matrices are approximately rank 1. In combination with the rank 1 tree level Yukawas, they result in Yukawa couplings that contain one very small eigenvalue; the corresponding eigenstates are identified with the first generation and
the obtained fermion mass spectrum is fully hierarchical.
Quark mixing is generically also very hierarchical. Mixing between the second and third as well as between the first and third generation is induced at the one loop level. Smallness of mixing between the first and second generation is guaranteed by the approximate rank 1 nature of the radiatively induced Yukawa matrix.

The qualitative picture agrees well with the observed SM fermion spectrum and quark mixing angles. At the quantitative level we find that our minimal framework can accommodate almost all observed flavor parameters. Starting with mixing angles: If the only source of flavor violation is the scalar soft masses the Cabibbo angle is found to be too small. However, $\lambda_C$  can easily be brought in agreement with experiment if trilinear couplings also violate flavor, to the degree that is allowed by vacuum stability. The additional flavor breaking adds an order one contribution to the up mass, which if dressed by a gluon can lead to a chromo-EDM. The EDM phenomenology of this scenario would thus be interesting to explore.

As for the mass spectrum, we found that the minimal framework predicts a correct electron-to-muon mass ratio, but the muon mass is too small by about an order of magnitude. We discussed various extensions of the minimal setup that allow to accommodate also the measured muon mass while keeping the electron-to-muon mass ratio fixed. 
One way to obtain a larger muon mass is to allow for hard SUSY breaking in the form of lepton-slepton-bino couplings that are larger than the hypercharge gauge coupling by a factor of $O(1)$. 
We also discussed an extended model with a gauge group enlarged by $U(1)_{B-L}$. Loops with the additional $B-L$ gauginos can give sizable contributions in particular to the lepton masses and bring the lepton spectrum in agreement with observation. 
The $B-L$ gauge coupling can be as large as $g_{B-L} \simeq 0.46$ at the PeV scale without spoiling gauge coupling unification.
Larger values of it can improve the agreement with the observed values, but at the cost of having a Landau pole before GUT scale. This  suggest  as interesting alternative to gauge $B-L$ to extend the MSSM with a non abelian gauge group. We defer this for a future work.


Frameworks that implement the idea of radiative fermion masses generically lead to spectra and mixing matrices that contain a minimum amount of hierarchy, given by a one loop factor. The observed masses and mixing angles in the neutrino sector show a pattern that is much less hierarchical with respect to the quarks and charged leptons. 
It would be interesting to see if simple extensions of our setups can accommodate also a realistic neutrino sector.
Studies in this direction are left for future work.

\begin{acknowledgments}
We would like to thank Nima Arkani-Hamed, Zackaria Chacko, Savas Dimopolous, Stefania Gori, Alex Kagan, Graham Kribs, Steve Martin, Mathias Neubert, Takemichi Okui, Eduardo Ponton, Lisa Randall and Jure Zupan for useful discussions and comments. 
WA would like to thank the Theoretical Physics 
Group at SLAC for hospitality and support.
This work was supported in part by National Science Foundation Grant No. PHYS-1066293 and the hospitality of the Aspen Center for Physics. Fermilab is operated by Fermi Research Alliance, LLC, under contract DE-AC02-07CH11359 with the United States Department of Energy. The research of WA was supported by the John Templeton Foundation. Research at Perimeter Institute is supported by the Government of Canada through Industry Canada and by the Province of Ontario through the Ministry of Economic Development \& Innovation. 
\end{acknowledgments}

$$
\href{https://www.youtube.com/watch?v=1JzrJZ0gIrA}{\bigstar} ~ 
\href{https://www.youtube.com/watch?v=O6m2WUyOTL0}{\bigstar} ~ 
\href{https://www.youtube.com/watch?v=omqtpZpjMMY}{\bigstar} ~ 
\href{https://www.youtube.com/watch?v=Yi0Dgu3ewtU}{\bigstar} \hspace{-52.8pt} 
\raisebox{2pt}{\textcolor{white}{\text{\fontsize{2}{3}\selectfont 54}} \hspace{4.7pt}
\textcolor{white}{\text{\fontsize{2}{3}\selectfont 74}} \hspace{4.7pt}
\textcolor{white}{\text{\fontsize{2}{3}\selectfont 90}} \hspace{4.6pt}
\textcolor{white}{\text{\fontsize{2}{3}\selectfont 14}}} $$

\begin{appendix}

\section{Observed Values of Fermion Masses and Mixings} \label{app:masses}

Here, we report the ratios of fermion masses that we compare to our model predictions throughout the paper. 
For definiteness we quote values for $\overline{\text{MS}}$ masses at the electro-weak scale, that we take to be $\hat \mu = m_t$, the $\overline{\rm MS}$ top mass at the scale of the top mass. In the case of the quark masses, we take into account 3-loop QCD running. In the case of the lepton masses we neglect the tiny QED corrections and use directly the pole masses. Using input from~\cite{Beringer:1900zz} we find
\begin{eqnarray} \label{eq:masses1}
 \frac{m_c}{m_t} = (3.7 \pm 0.1) \cdot 10^{-3} &,& \frac{m_u}{m_t} = (8.1 \pm 2.0) \cdot 10^{-6} ~,~ \frac{m_u}{m_c} = (2.2 \pm 0.6) \cdot 10^{-3} ~,\\[4pt] \label{eq:masses2}
 \frac{m_s}{m_b} = (1.9 \pm 0.1)\cdot 10^{-2} &,& \frac{m_d}{m_b} = (9.9 \pm 0.8)\cdot 10^{-4}~,~ \frac{m_d}{m_s} = (5.2 \pm 0.5)\cdot 10^{-2}~,\\[4pt] \label{eq:masses3}
 \frac{m_\mu}{m_\tau} = 5.95 \cdot 10^{-2} \qquad~~ &,& \frac{m_e}{m_\tau} = 2.88 \cdot 10^{-3}~,~ \qquad~~ \frac{m_e}{m_\mu} = 4.84 \cdot 10^{-3}~.
\end{eqnarray}
We do not quote the extremely small uncertainties on the lepton mass ratios, as they are of no relevance for our study.

The measured values of the CKM parameters at the electro-weak scale are~\cite{Beringer:1900zz}
\begin{eqnarray}
\lambda_C &\simeq& 0.225 ~, \\
 |V_{cb}| &=& (4.09 \pm 0.11) \times 10^{-2} ~, \\ 
 |V_{ub}| &=& (4.15 \pm 0.49)\times 10^{-3} ~.
\end{eqnarray}

\section{Vacuum Stability Constraints in the Presence of Flavor Violation}\label{app:vac}

In this appendix we discuss vacuum stability constraints on the sfermion trilinear couplings and the Higgsino mass in the presence of generic flavor mixing in the sfermion soft masses. 
We focus on the case of absolute stability. The analysis of meta-stability bounds, where one allows the electro-weak vacuum to decay at time scales larger than the age of the universe, is beyond the scope of this work.

We consider the case of sufficiently large $\tan\beta$ such that we can neglect the effect of the down-type Higgs $H_d$. For simplicity we will also consider only real mass insertions.
In order to obtain a bound on the stop trilinear coupling in the presence of $O(1)$ squark mixing, we consider the scalar potential of the MSSM in the most general D-flat direction containing the up-type Higgs $H_u$ as well as all 6 up-type quarks. We find the following bound on the stop trilinear
 \begin{equation} \label{eq:flavoredAtbound}
 |A_t|^2 \lesssim \left( 3 + t_L^2 + t_R^2 \right) ( \bar m_{\tilde t_L}^2 + \bar m_{\tilde t_R}^2) ~,
\end{equation}
where the masses $\bar m_{\tilde t_L}^2$ and $\bar m_{\tilde t_R}^2$ are given by
\begin{eqnarray} \label{eq:mt}
\bar m_{\tilde t_L}^2 &=& c_L^2 (m_{\tilde u_L}^2)_{33} + 2 c_L s_L \tilde c_L (m_{\tilde u_L}^2)_{23} + s_L^2 ( \tilde s_L^2 (m_{\tilde u_L}^2)_{11} + \tilde c_L^2 (m_{\tilde u_L}^2)_{22} + 2 \tilde s_L \tilde c_L (m_{\tilde u_L}^2)_{12} ) ~, \\   \label{eq:mt2}
\bar m_{\tilde t_R}^2 &=& c_R^2 (m_{\tilde u_R}^2)_{33} + 2 c_R s_R \tilde c_R (m_{\tilde u_R}^2)_{23} + s_R^2 ( \tilde s_R^2 (m_{\tilde u_R}^2)_{11} + \tilde c_R^2 (m_{\tilde u_R}^2)_{22} + 2 \tilde s_R \tilde c_R (m_{\tilde u_R}^2)_{12} ) ~,
\end{eqnarray}
and we introduced the shorthand notation
\begin{eqnarray}
 t_{L/R} = \tan\theta_{L/R} ~&,&~~ c_{L/R} = \cos\theta_{L/R} ~,~~ s_{L/R} = \sin\theta_{L/R} ~, \\
 \tilde t_{L/R} = \tan\tilde\theta_{L/R} ~&,&~~ \tilde c_{L/R} = \cos\tilde\theta_{L/R} ~,~~ \tilde s_{L/R} = \sin\tilde\theta_{L/R} ~.
\end{eqnarray}
The inequality in~(\ref{eq:flavoredAtbound}) has to hold for all choices of the angles $\theta_{L/R}, \tilde\theta_{L/R}$. Equivalently, one can impose the bound with the right-hand side minimized with respect to the angles $\theta_{L/R}, \tilde\theta_{L/R}$. For $\theta_{L/R} = \tilde \theta_{L/R} = 0$ one recovers the bound in~(\ref{eq:Atbound}). In general,~(\ref{eq:flavoredAtbound}) can be stronger than~(\ref{eq:Atbound}).

In order to obtain bounds on the Higgsino mass in the presence of $O(1)$ squark and slepton mixing, we have to consider the scalar potential of the MSSM in directions containing the up-type Higgs $H_u$ as well as all 6 down-type quarks or leptons. As there is no exact D-flat direction in this case, for any given field direction, the derived bounds are only necessary but usually not sufficient to guarantee the absence of charge and color breaking minima. Analogously to section~\ref{sec:stability} we chose exemplarily $|\tilde \ell_L| = |\tilde \ell_R| = H_u/\sqrt{2}$ and $|\tilde d_L| = |\tilde d_R| = H_u/\sqrt{2}$. In this case we find the necessary conditions
\begin{eqnarray} \label{eq:flavoredmubound}
 |\mu|^2 &\lesssim& (\bar m_{\tilde \tau_L}^2 + \bar m_{\tilde \tau_R}^2) \left[ \frac{1}{2} + \frac{9}{32} \frac{v^2}{m_\tau^2} \left|\frac{1}{t_\beta} + \epsilon_\tau \right|^2( x_{\tau1} g_1^2 + x_{\tau2}g_2^2) \right]~, \\ \label{eq:flavoredmubound2}
 |\mu|^2 &\lesssim& (\bar m_{\tilde b_L}^2 + \bar m_{\tilde b_R}^2) \left[ \frac{1}{2} + \frac{9}{32} \frac{v^2}{m_b^2} \left|\frac{1}{t_\beta} + \epsilon_b \right|^2(x_{b1} g_1^2 + x_{b2}g_2^2) \right]~.
\end{eqnarray}
The masses $\bar m_{\tilde \tau_L}^2$, $\bar m_{\tilde \tau_R}^2$, $\bar m_{\tilde b_L}^2$, and $\bar m_{\tilde b_R}^2$ are defined analogously to the up-type masses in~(\ref{eq:mt}) and~(\ref{eq:mt2}). The coefficients $x_i$ are given by
\begin{eqnarray}
 x_{\tau1} &=& 1 + \frac{4}{3} t_L^2 + \frac{4}{9} t_L^2 t_R^2 + \frac{4}{9} t_L^2 c_R^2 - \frac{4}{9} s_R^2 + \frac{1}{9} t_R^2 c_L^2 - \frac{1}{9} s_L^2 ~,\\
 x_{b1} &=& 1 + \frac{16}{27} t_L^2+ \frac{20}{27} t_R^2 + \frac{4}{9} t_L^2 t_R^2 + \frac{1}{81} c_L^2 t_R^2 - \frac{1}{81} s_L^2 + \frac{4}{81} c_R^2 t_L^2 - \frac{4}{81} s_R^2  ~,\\
 x_{\tau2} = x_{b2} &=& 1 + \frac{4}{9} t_L^2+ \frac{8}{9} t_R^2 + \frac{4}{9} t_L^2 t_R^2 + \frac{1}{9} c_L^2 t_R^2 - \frac{1}{9} s_L^2  ~.
\end{eqnarray}
The inequalities in~(\ref{eq:flavoredmubound}) and~(\ref{eq:flavoredmubound2}) have to hold for all choices of the angles $\theta_{L/R}, \tilde\theta_{L/R}$. In general, (\ref{eq:flavoredmubound}) and~(\ref{eq:flavoredmubound2}) can be stronger than~(\ref{eq:mubound}) and~(\ref{eq:mubound2}).

\section{Flavor and EDM Constraints} \label{app:flavor}

The radiative mechanism of fermion mass generation discussed in this work requires $O(1)$ flavor mixing in the sfermion soft masses. Such flavor mixing is subject to strong constraints from low energy flavor observables~\cite{Gabbiani:1996hi,Altmannshofer:2009ne,Altmannshofer:2013lfa}.
In particular CP violation in Kaon mixing, given by the observable $\epsilon_K$, is exceedingly sensitive to the mixing between the first two generations of down-type squarks. In order not to spoil the reasonable agreement between the experimental results and the SM prediction for $\epsilon_K$, we find that the SUSY scale $\tilde m$ has to be of the order of few PeV. We expect this to be the strongest bound on the SUSY breaking scale in our model. 

In the presence of large sfermion mixing it is known that also electric dipole moments (EDMs) are generically very sensitive low energy probes of heavy supersymmetric particles. Current experimental bounds on EDMs of the neutron and electron can probe squarks and sleptons with masses of O(100)~TeV~\cite{McKeen:2013dma,Altmannshofer:2013lfa}, and prospects for further experimental improvements are excellent. However, in the framework of radiative fermion masses, there is the potential of an approximate phase alignment between the light fermion masses and the corresponding dipole moments~\cite{Borzumati:1999sp}. This is because the leading contributions to the dipole come from dressing the leading 1-loop diagram for the mass by a gluon or a photon. However, in some of our models, the mass of the first generation fermions came from more than one diagram. In particular, once flavor violating trilinears were introduced in section~\ref{sec:Aterms}, the up mass received comparable contributions from soft masses and $A$-terms, as seen in equation~(\ref{eq:up_mass}). In this case we expect the mass and the dipole to be misaligned. A full exploration of the EDM phenomenology is beyond the scope of our work.

\section{The Higgs Mass} \label{app:higgs}

\begin{figure}[t]
\centering
\includegraphics[width=0.8\textwidth]{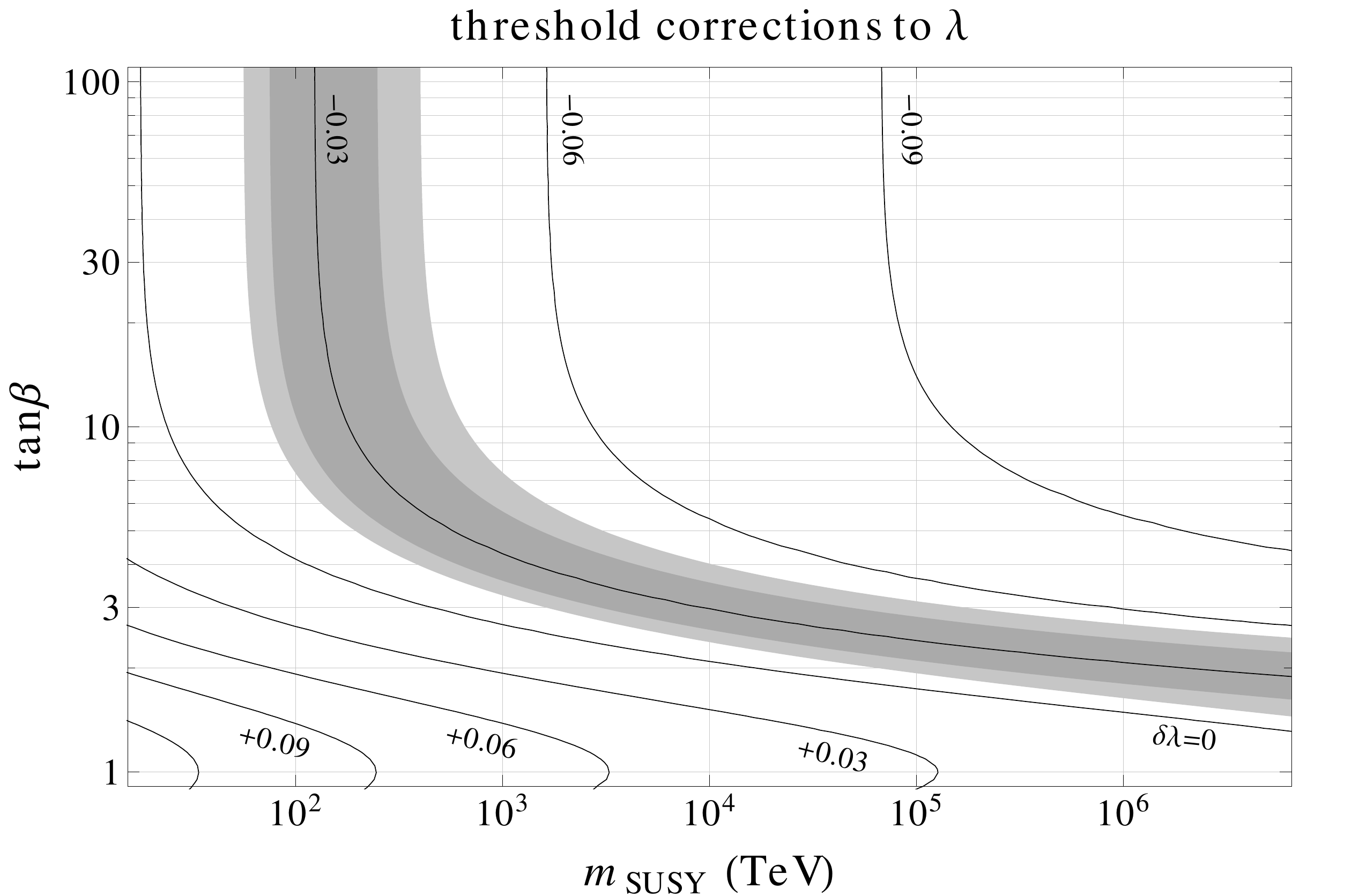}
\caption{Threshold corrections to the Higgs quartic coupling at the SUSY scale that are required to obtain a Higgs mass of $m_h \simeq 125.5$~GeV. The gray bands exemplarily indicate the $1\sigma$ and $2\sigma$ uncertainties coming from the top mass, strong gauge coupling, and the Higgs mass.}
\label{fig:Mhiggs}
\end{figure}

As squarks are very heavy in the discussed framework (at least few PeV in order to avoid constraints from Kaon mixing), the SM-like Higgs is predicted to be somewhat heavier than the observed value of $m_h \simeq 125.5$~GeV. In the high scale SUSY scenario that we are working in, the observed Higgs mass is reproduced for a SUSY scale of around 10 TeV if $\tan\beta \gtrsim 10$~\cite{Giudice:2011cg,Draper:2013oza,Bagnaschi:2014rsa}. In figure~\ref{fig:Mhiggs} we show in the plane of the SUSY scale and $\tan\beta$ the threshold corrections $\delta \lambda$ to the Higgs quartic coupling that are required at the SUSY scale in order to obtain a Higgs mass of 125.5 GeV. At the SUSY scale we set the Higgs quartic to
\begin{equation}
\lambda(m_\text{SUSY}) = \frac{1}{4} \Big[ g_1^2(m_\text{SUSY}) + g_2^2(m_\text{SUSY}) \Big] \cos^2 2\beta + \delta \lambda ~.
\end{equation}
We use SM 2-loop RGEs~\cite{Machacek:1983tz,Machacek:1983fi,Machacek:1984zw,Luo:2002ey} for the evolution of the Higgs quartic between the SUSY scale and the electro-weak scale, taking into account contributions from all gauge couplings and the top Yukawa. We incorporate 2-loop threshold corrections at the electro-weak scale from~\cite{Buttazzo:2013uya}. 
The gray band in the plot of figure~\ref{fig:Mhiggs} exemplarily shows the $1\sigma$ and $2\sigma$ uncertainty on the required threshold corrections at the SUSY scale, taking into account the uncertainty on the top quark pole mass $m_t = 173.34 \pm 0.76$~GeV~\cite{ATLAS:2014wva}, the strong gauge coupling $\alpha_s(m_Z) = 0.1185 \pm 0.0006$~\cite{Beringer:1900zz}, and the Higgs mass $m_h = 125.6 \pm 0.3$~GeV\footnote{We obtain this value from a naive weighted average of the ATLAS result $m_h = 125.5 \pm 0.2 ^{+0.5}_{-0.6}$~GeV~\cite{Aad:2013wqa} and the CMS result $m_h = 125.7 \pm 0.3 \pm 0.3$~GeV~\cite{CMS-PAS-HIG-13-005}.}.

For $\tan\beta \gtrsim 10$, the required threshold corrections are at the level of $\delta \lambda \simeq -0.05$ for a PeV scale spectrum.
Note that sizable stop trilinear couplings $|A_t| \sim m_{\tilde t}$ typically lead to positive threshold corrections of the order of $\delta \lambda \sim + \text{few} \times 10^{-2}$. A very large stop trilinear $|A_t| \gtrsim \sqrt{12} m_{\tilde t}$ would lead to negative corrections. However, such a large $A_t$ is in conflict with vacuum stability constraints, see~(\ref{eq:Atbound}).   

Various viable mechanisms exist that lead to negative threshold corrections. One example is to add a singlet $S$ with a SUSY mass $M_S$ that couples to the MSSM Higgs bosons
\begin{equation}
 W_S \supset \frac{1}{2} M_S S^2 + \lambda_S S H_u H_d ~.
\end{equation}
The additional scalar modifies the Higgs potential and leads to a shift in the Higgs quartic. We find
\begin{equation}
 \delta \lambda \simeq \frac{2 \lambda_S^2}{M_S^2 + m_S^2} \left( \frac{m_S^2}{t_\beta^2} -\frac{2 \mu M_S}{t_\beta} - \mu^2 \right) ~,
\end{equation}
where $m_S^2$ is the soft mass squared for the scalar component of~$S$ and we expanded in large~$t_\beta$. The first term in parenthesis proportional to the soft mass squared gives a positive definite contribution to the Higgs quartic, but it is completely negligible in the large $\tan\beta$ regime. The remaining terms are proportional to the supersymmetric masses $M_S$ and $\mu$ and can reduce the Higgs quartic.
As long as $M_S$ is not too heavy, moderate $\lambda_S$ can easily lead to threshold corrections of the required size.

Another  avenue to lower the  tree level Higgs mass is a hard SUSY breaking negative quartic coupling. As discussed in section~\ref{sec:hard}, hard SUSY breaking can help in raising the muon mass in our framework. The only hard SUSY breaking operator that enters at order $F/M^2$ in the MSSM is~\cite{Martin:1999hc} 
\begin{equation}
\int d^2\theta \frac{X}{M^2}(H_u H_d)^2=\frac{F}{M^2}(H_u H_d)^2\,.
\end{equation}
This operator however leads to a $\tan\beta$ suppressed quartic. To lower the Higgs mass in our framework we need to write 
\begin{equation}
\int d^4\theta \frac{X^\dagger X}{M^4}(H_u^\dagger H_u)^2 = \frac{F^2}{M^4}|H_u|^4\,.
\end{equation}
The value of $F^2/M^4$ needed to increase the muon mass (of order $g_1$) is sufficient to lower the Higss mass to its observed value. 

Finally, we note that another option is to  make the gauginos Dirac fermions~\cite{Fox:2002bu}. Notice that a sizeable Majorana mass for the gauginos is a necessary ingredient for our mechanism to work, therefore one would need same order Majorana and Dirac fermion masses for the gauginos. 

\end{appendix}


\end{document}